\documentclass{article}

\RequirePackage{amsthm,amsmath,amsfonts,amssymb,amsthm}
\RequirePackage[urlcolor=blue]{hyperref}
\RequirePackage{graphicx, tikz, pgfplots}
\RequirePackage[ruled,vlined]{algorithm2e}
\RequirePackage{appendix}
\usepackage{natbib, ulem, authblk, float, setspace, booktabs}
\usepackage[margin=1in]{geometry}

\definecolor{pptblue}{RGB}{68, 114, 196}
\definecolor{pptred}{RGB}{210, 0, 0}

\SetKwInput{KwInputs}{Inputs}
\SetKwInput{KwOutput}{Output}

\pgfplotsset{compat=1.17}

\title{Factorized Binary Search: change point detection in the network structure of multivariate high-dimensional time series}

\author[1]{Martin Ondrus}
\author[2]{Emily Olds}
\author[1,2]{Ivor Cribben}
\affil[1]{Neuroscience and Mental Health Institute, University of Alberta, Canada}
\affil[2]{Alberta School of Business, University of Alberta, Canada}

\begin{document}

\maketitle{}

\begin{abstract}
Functional magnetic resonance imaging (fMRI) time series data presents a unique opportunity to understand the behavior of temporal brain connectivity, and models that uncover the complex dynamic workings of this organ are of keen interest in neuroscience.  We are motivated to develop accurate change point detection and network estimation techniques for high-dimensional whole-brain fMRI data.  To this end, we introduce \textit{factorized binary search} (FaBiSearch), a novel change point detection method in the network structure of multivariate high-dimensional time series in order to understand the large-scale characterizations and dynamics of the brain. FaBiSearch employs non-negative matrix factorization, an unsupervised dimension reduction technique, and a new binary search algorithm to identify multiple change points. In addition, we propose a new method for network estimation for data between change points.  We seek to understand the dynamic mechanism of the brain, particularly for two fMRI data sets. The first is a resting-state fMRI experiment, where subjects are scanned over three visits.  The second is a task-based fMRI experiment, where subjects read Chapter 9 of \textit{Harry Potter and the Sorcerer's Stone}.  For the resting-state data set, we examine the test-retest behavior of dynamic functional connectivity, while for the task-based data set, we explore network dynamics during the reading and whether change points across subjects coincide with key plot twists in the story.  Further, we identify hub nodes in the brain network and examine their dynamic behavior.  Finally, we make all the methods discussed available in the R package \textbf{fabisearch} on CRAN.
\end{abstract}


\doublespacing

\section{Introduction}
Functional magnetic resonance imaging (fMRI) is widely used as an indirect method of measuring brain activity using the blood-oxygen-level-dependent (BOLD) signal \citep{ogawa}. With increases in neuronal activity, glucose and oxygen must be delivered to the neurons from the blood stream causing a local increase in oxyhemoglobin. This subsequently increases the ratio of oxyhemoglobin to deoxyhemoglobin, and provides the basis of measurement of the BOLD signal in fMRI. Typically, multiple slices of the subject's brain are scanned and then divided into small (a few millimeters in dimension) cubes called voxels. By taking multiple scans of each slice sequentially over time, a time series of voxel activity can be created.

With the advent of ever increasing computational capacity, and general interest in scalable, data-driven techniques for characterizing neurological phenomena, fMRI data presents a unique opportunity. By studying the relationships among voxels (or a cluster of voxels often referred to as regions of interest, or ROIs), we can better characterize brain connectivity leading to a greater understanding of the brain and extensive clinical implications \citep{bullmore, sporns}. Functional connectivity (FC) seeks to define the relationships \citep{biswal} by means of correlation, covariance, and precision matrices amongst other techniques (see \cite{cribbenfiecas} for a review). Graph theory remains at the core of most FC based estimation methods, and for good reason; it intuitively represents the brain as a graphical model with ROIs as nodes and temporal dependence as edges. These FC relationships have vast scientific and clinical implications. Not only can FC be used to understand brain processes during tasks or resting, but also to characterize clinical diagnoses. For example, schizophrenia has long been associated with abnormal FC, characterized by disorder in network patterns such as small-worldness \citep{B.Nejad2012}.

However, these FC methods typically assume that the FC between the ROIs remains constant throughout the duration of the experiment. While this reduces computational time and complexity, it fails to capture the evolving and dynamic behavior of the brain. Evidence of a changing, non-stationary FC over the experiment has been demonstrated in both resting-state data \citep{delamillieure, doucet} and task-based experiments \citep{fox, debener, eichele, cribben3}. Although some neural processes are understood through the interaction of brain regions, many complex cognitive processes are not well characterized. In most cases, the circuitry is assumed to remain static, not allowing for changes to be reflected in each individual as the task itself evolves, such as when the narrative of a story develops during a reading task. \cite{Varela2001} suggests that neural assemblages respond to internal and external stimuli through changes in synchronization.  Consequently, it is important to not only study how different brain regions synchronize in activity over time, but also how this synchronization evolves throughout the time course and as a result of different stimuli \citep{Gonzalez-Castillo2018}.

Numerous statistical methods have emerged to capture and express the dynamic nature of FC. First, moving window approaches were introduced that extend covariance, correlation, or precision matrix methods into a time-varying context. These approaches define sequential blocks of time points and estimate FC within each block. By estimating FC across blocks from the beginning to the end of the time course, time-varying FC can be estimated (see \cite{hutchison} for a review).  Although the moving-window approach is a computationally practical way of determining FC, it has limitations \citep{hutchison}. For one, the resulting FC patterns are heavily influenced by the choice of block size, which can result in vastly different FC patterns. Additionally, this technique gives no weight to time points outside of those included in the window.

Instead of a moving window, change point methods have also been applied to this problem.  Here, the objective is to find the optimal windows for stationary structures. There exists an extensive literature and a long history on change points beginning with \cite{page}.  The most widely discussed problems have been concerned with finding change points in univariate time series and more recently, with detecting multiple change points in multivariate time series.  For example, \cite{aue2009} introduced a method to detect changes in the covariance matrix of a multivariate time series, \cite{dette2016} proposed a test where the dimension of the data is fixed, while more recently \cite{kao} considered the case where the dimension of the data increases with the sample size (they also investigated change point analysis based on principal component analysis). \cite{sundararajan} proposed a new method for detecting multiple change points in the covariance structure of a multivariate piecewise-stationary process.  Using a combination of principal components analysis and wavelets to transform the time series, \cite{ChoFryzlewicz2015} segmented the multivariate time series into partitions based on the second-order structure. 

There have also been many new statistical methods developed specifically for an application to neuroimaging data. \cite{cribben1} first introduced a change point method for estimating dynamic functional connectivity by considering change points in precision matrices (undirected graphs) using binary segmentation and the bayesian information criterion metric. Accordingly, \cite{cribben2, schroder,  kirch,Gibberd2014,avanesov,dai2019,anastasiou} then proposed further methods for estimating FC change points. \cite{Koepcke2016,mosqueiro,Xiao2019} also considered change points in spike trains.  While these methods are effective, they all are limited in the number of time series that can be considered. Subsequently, there has been a drive to extend techniques to high-dimensional spaces, specifically the case $p >> T$, where $p$ is the number time series and $T$ is the total length of the time series.  To this end, \cite{Cribben2017} introduced the network change point detection method, which uses both change point and community detection techniques to estimate change points by examining the time evolving community network structure of multivariate time series.  In addition, \cite{ofori2019} introduced a new method that firstly presents each network snapshot of fMRI data as a linear object and finds its respective univariate characterization via local and global network topological summaries and then adopts a change point detection method for (weakly) dependent time series based on efficient scores. While these methods are an adequate starting point to understand dynamic neural processes, the true granularity in which the brain functions is lost with less comprehensive cortical maps of the brain. In practice, these cortical maps define fewer ROIs for the same size of cerebrum, and so activity is averaged across a larger surface area. This decreases the specificity of individual ROI activity and effectively assumes that activity is the same within each, which is highly unlikely given the complexity of most cognitive processes.

As such, we are motivated to develop accurate change point detection and network estimation techniques for high-dimensional whole-brain fMRI data. We seek to understand the dynamic mechanism of the brain through these techniques, particularly for two experiments using the \cite{Gordon2016} ROI atlas ($p=333$).  The first data set is a test-retest resting-state fMRI experiment of 25 participants scanned over three visits.  Here, we examine the change in an individual's brain dynamics behavior over time within a scanning session, the test-retest reliability of dynamic functional connectivity (across scans or longitudinal measurements), and whether commonalities exist across subjects' estimated networks or functional states.  The second data set is a task-based fMRI experiment of 8 subjects reading a chapter of \textit{Harry Potter and the Sorcerer's Stone} \citep{Rowling2012}. Again, using a large number of brain regions ($p=333$), we seek to understand the behavior of whole-brain network dynamics, but crucially, we hope to understand how the change points and estimated networks coincide with key features in the plot.

To this end, we introduce a new method, called \textit{factorized binary search} (FaBiSearch), to detect multiple change points in the network (or clustering) structure of multivariate high-dimensional time series. FaBiSearch has the following unique and important attributes. First, FaBiSearch is, to the best of our knowledge, the first statistical method to use non-negative matrix factorization (NMF) for finding change points in the network (or clustering structure) in multivariate high-dimensional time series, which allows us to understand the large-scale behavior and dynamics of the brain. NMF is an unsupervised method of dimension reduction and clustering commonly used in text analysis. We have many motivations for using NMF: it has an inherent clustering (or community) property, where it automatically clusters the time series, which is important in neuroimaging. Additionally, compared to other clustering techniques, NMF does not have any constraints on the interaction of variables to be orthogonal, as in Principal Component Analysis (PCA) \citep{karlpearson} or independent, as in Independent Component Analysis (ICA) \citep{jutten1991blind}. Such constraints do not necessarily hold in fMRI data or multivariate time series in general. Consequently, NMF is more flexible in modelling the interactions between variables and allows for some overlap in the basis components \citep{lee1999learning}. Furthermore, because of the non-negative constraint of the data and factors, the learned associations are non-negative as well. This is greatly beneficial in an fMRI context where anticorrelations between brain regions are not easily interpretable. Second, FaBiSearch is suitable for detection of multiple change points, which are common not only in task-based fMRI experiments, but also in resting-state experiments.  Third, FaBiSearch is scalable ($p>>T$) in that it is not limited by the dimensionality of the problem space and is therefore ideal for characterizing large, changing network structures, such as those in the brain.  For the simulated data sets we consider in this work, we find that FaBiSearch has a superior performance to previous state-of-the-art methods. In addition, for simulations where the subject alternates between two states (such as task-based fMRI data) and where the subject transitions between states, FaBiSearch performs very well and clearly outperforms the other methods.  Fourth, for FaBiSearch, we introduce a novel binary search algorithm to identify multiple change points in a high-dimensional setting, which are common not only in task-based fMRI experiments, but in resting-state experiments.  This new search algorithm dramatically increases computational efficiency. Fifth, the NMF element of FaBiSearch allows us to define a new method for the estimation of networks for data between change points, which provides a visual display of the clustering structure and the FC networks between the brain regions. Sixth, while motivated by fMRI data, FaBiSearch may also be applicable to electroencephalography (EEG), magentoencephalography (MEG) and electrocorticography (ECoG) data, and other multivariate high-dimensional time series applications where the network, community or clustering structure is changing over time.  Finally, the R package \textbf{fabisearch} implementing the methods is available from CRAN \citep{fabisearch}.

The setup of this paper is as follows. In Section \ref{sec:prelim} we define notation for the method, while in Section \ref{sec:prob}, we describe the problem setting and goal of our approach. In Section \ref{sec:methods}, we describe our novel method, FaBiSearch, and in Section \ref{sec:netest} the method for estimating networks between pairs of change points. In Sections \ref{sec:data}, we describe the simulated and fMRI data sets and we outline the results in Section \ref{sec:results}. We have a discussion in Section \ref{sec:discussion} before concluding in Section \ref{sec:conclusion}.

\section{Preliminaries}\label{sec:prelim}

An entry in the $i$th row and $j$th column of a matrix $\boldsymbol{A}$ is denoted by $\boldsymbol{A}_{ij}$. A matrix is non-negative, $\boldsymbol{A} \geq 0$, if and only if all entries of the matrix are non-negative, that is, $\boldsymbol{A}_{ij} \geq 0$ $\forall i,j$. For a matrix of time series data $\boldsymbol{X} \in \mathbb{R}^{T \times p}$, $T$ is total the number of discrete time points and $p$ is the number of variables. The set of $k$ number of \textit{true} change points in $\boldsymbol{X}$ are denoted by $\tau = \{\tau_1, ..., \tau_k\}$ where $1 < \tau_1 < \tau_2 ... \tau_k < T$. For some pair of time points $t_1, t_2 < T$ where $t_1 < t_2$, we denote the corresponding range of samples in $\boldsymbol{X}$ as $\boldsymbol{X}_{t_1:t_2,}$. Similarly, for some pair of variable indices $p_1, p_2 < p$ where $p_1 < p_2$, we denote the corresponding range of columns in $\boldsymbol{X}$ as $\boldsymbol{X}_{,p_1:p_2}$. We denote the cardinality of a set with $|~~|$ (e.g., $|T|$). We define $\delta$ as the minimum distance between two change points, which is equivalent to the minimum sample size in our setup. We refer to the set of all ordered (from smallest to largest) possible change points as $Q = \{1 + \delta,..., T - \delta\}$, $m$ candidate change points as $\hat{Q} = \{\hat{q}_1,...,\hat{q}_m\}$, and $s$ change points as $\hat{Q}^* = \{\hat{q}^*_1,...,\hat{q}^*_s\}$, where $\hat{Q}^* \subseteq \hat{Q} \subseteq Q \subseteq [1,...,T]$. 

\section{Problem Setting}\label{sec:prob}

Consider a high-dimensional, multivariate time series $\boldsymbol{X} = \{ x_t \in \mathbb{R}^p : t = 1, 2, ... , T \}$ where $x$ is a vector of $p$ variables at time $t$, and $p >> T$. Further, assume that $\boldsymbol{X}$ is piece-wise stationary, with structural breaks defined by $\tau$. In the case that $\boldsymbol{X}$ does not have change points, $k = 0$ and $\tau$ is an empty set, and $\boldsymbol{X}$ is stationary. If $\boldsymbol{X}$ is non-stationary, we can segment $\boldsymbol{X}$ into $k + 1$ stationary segments $\boldsymbol{S}_i$ defined by $\tau$, such that $\boldsymbol{S}_i = \{x_t \in \mathbb{R}^p : \tau_{i-1} < t \leq \tau_i\}$. Our goal in this setting is that, given $\boldsymbol{X}$, we are interested in recovering $\tau$ without any information on the location or number of true change points.

\section{Change Point Model Setup}\label{sec:methods}

In this section, we introduce our Factorized Binary Search (FaBiSearch) method. To do so, we first describe non-negative matrix factorization (NMF), then our novel segmentation method, and finally the permutation test procedure. Our method is an unsupervised, offline \citep{aminikhanghahi2017survey}, and test-based approach for change point detection. A test-based approach attempts to first find candidate change points through partitioning of the time series, and then evaluate the candidates through permutation testing \citep{antoch2001permutation}, which has been extensively applied in the literature \citep{james1987tests, andrews1993tests, shao2010testing, cribben1, xiong, chen2015graph, anastasiou2022detecting}.

\subsection{Non-negative matrix factorization}

NMF is a matrix factorization technique which constrains the input matrix and subsequent factors to non-negative values \citep{NIPS2000_1861}. It is commonly used as an unsupervised dimension reduction method and allows for high-dimensional data sets to be projected into simpler and ultimately more workable dimensions.  Our motivation for using NMF is that it has an inherent clustering (or community) property, where it automatically clusters the time series, which is important in neuroimaging where it is of interest to find communities (or functional states) that correspond to closely knitted groups of time series (or nodes).

\begin{figure}[ht]
\begin{center}
  \includegraphics[width=0.5\linewidth]{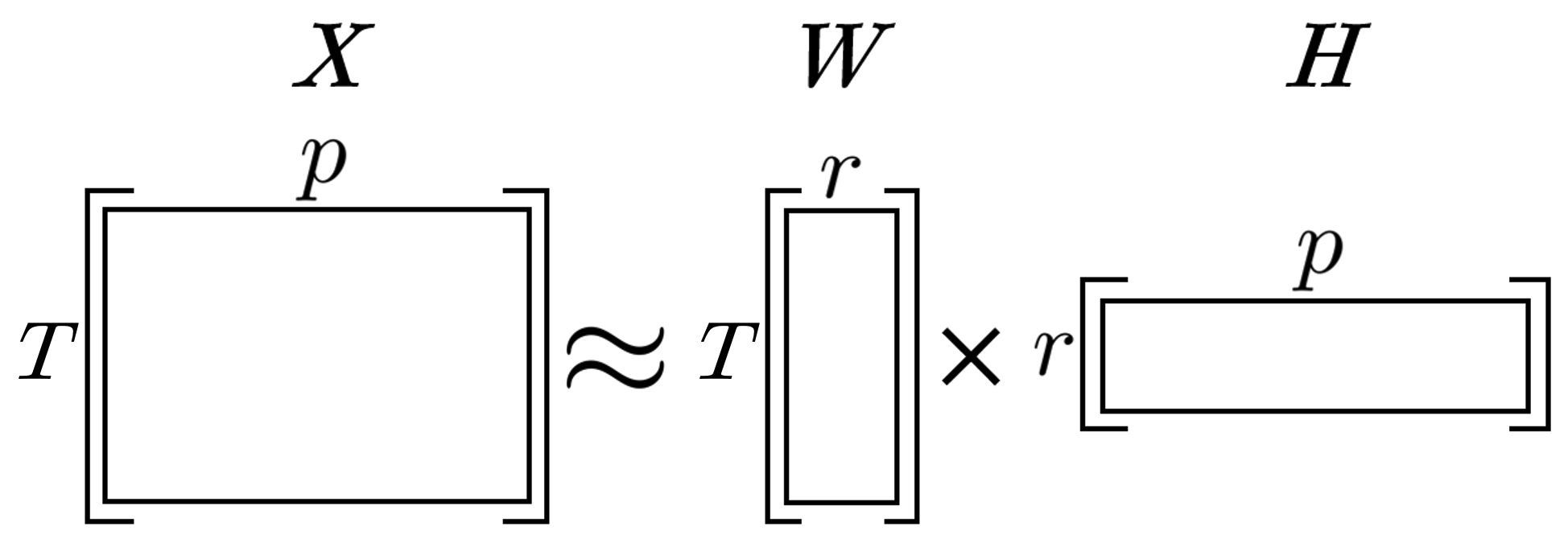}
  \caption{Graphic of non-negative matrix factorization (NMF), where $\boldsymbol{X} \in \mathbb{R}^{T \times p}$ is approximated by the product of low dimensional factors $\boldsymbol{W} \in \mathbb{R}^{T \times r}$ and $\boldsymbol{H} \in \mathbb{R}^{r \times p}$, where $r << min(T,p)$.}
  \label{fig:NMFgeneral}
\end{center}
\end{figure}

More formally, consider a non-negative matrix of $T$ samples and $p$ variables, $\boldsymbol{X} \geq 0 \in \mathbb{R}^{T \times p}$. NMF seeks to minimize the distance between the original matrix $\boldsymbol{X}$, and the product of two low rank factors, a coefficient matrix, $\boldsymbol{W} \geq 0 \in \mathbb{R}^{T \times r}$, and a basis matrix, $\boldsymbol{H} \geq 0 \in \mathbb{R}^{r \times p}$ with rank $r \in \mathbb{N} << min(T, p)$ (Figure \ref{fig:NMFgeneral}), such that $\boldsymbol{X} \approx \boldsymbol{W H}$. The interpretation of rank is context dependent, however, due to the clustering property of NMF \citep{Li2014, Li2004}, it can be seen as the number of unique clusters in $\boldsymbol{X}$. The value of $r$ is chosen \textit{a priori} and should be large enough that $\boldsymbol{W} \boldsymbol{H}$ retains the key information of $\boldsymbol{X}$ while also small enough that the addition of noise has a minimal effect; choosing $r$ is key for managing the bias-variance trade-off.

We utilize a loss measure based on the generalized Kullback-Leibler divergence from \cite{NIPS2000_1861} (or \textit{I}-divergence: \citealp{honkela}) to assess model fit,
\begin{equation}
\text{KLD}(\boldsymbol{X}||\boldsymbol{W}\boldsymbol{H}) = \sum_{\substack{i,j}}(\boldsymbol{X}_{ij} \log \frac{\boldsymbol{X}_{ij}}{(\boldsymbol{WH})_{ij}} - \boldsymbol{X}_{ij} + (\boldsymbol{WH})_{ij}),
\label{eqn:KD}
\end{equation}
\noindent where the asymmetric divergence between the input matrix $\boldsymbol{X}$ and the NMF factorization $\boldsymbol{W} \boldsymbol{H}$ is calculated for each entry in the $i$th row and $j$th column. In addition to this loss metric, \cite{NIPS2000_1861} also describe other loss metrics and algorithms utilizing multiplicative updates to find factors $\boldsymbol{W}$ and $\boldsymbol{H}$ given an input matrix $\boldsymbol{X}$. Given NMF is non-convex \citep{NIPS2000_1861},  NMF is typically run with multiple random initializations, which we denote as $n_{run}$, to help convergence.

As mentioned previously, rank $r$ is a key parameter in NMF, however, in many cases it is unknown. Thus, to find the optimal rank, we provide a data-driven solution for its estimation, which we define as \texttt{optrank}, which has been adapted from \cite{Frigyesi2008}. Let $\boldsymbol{X'}$ be the matrix resulting from randomly permuting the columns and rows of $\boldsymbol{X}$. Furthermore, let $\boldsymbol{WH}_{r}$ and $\boldsymbol{WH'}_{r}$ be the estimated matrices using NMF on $\boldsymbol{X}$ and $\boldsymbol{X'}$ respectively at some rank $r$. The approach compares the improvement in loss from increasing rank in $\boldsymbol{X}$ and $\boldsymbol{X'}$, and stops when the improvement in loss is no better than noise. More formally, starting at $r = 1$, calculate $\Delta_{X} \gets \text{KLD}(\boldsymbol{X}||\boldsymbol{WH}_{r+1}) - \text{KLD}(\boldsymbol{X}||\boldsymbol{WH}_{r})$ and $\Delta_{X'} \gets \text{KLD}(\boldsymbol{X'}||\boldsymbol{WH}_{r+1}) - \text{KLD}(\boldsymbol{X'}||\boldsymbol{WH}_{r})$. Iterate through $r$ while $\Delta_{X} > \Delta_{X'}$, and then the optimal rank, $r^*$, is the first $r$ where $\Delta_{X} < \Delta_{X'}$.

\subsection{Segmentation using binary search}

The most commonly used segmentation method for change point detection is binary segmentation \citep{Douglas1973ALGORITHMSFT, RAMER1972244, doi:10.1086/620282}, where a univariate collection of samples is sequentially evaluated across the time index set. This can then be extended to the multivariate time series setting, by choosing some representative criterion which we can sequentially evaluate across the time index set. In particular, for data $\boldsymbol{X} \in \mathbb{R}^{T \times p}$, binary segmentation proceeds such that each time point $t \in [1...T]$ is evaluated sequentially for some chosen model measure, $L$ (e.g., loss, likelihood, or some information criterion). The time point point which minimizes or maximizes the measure is picked as the candidate change point, $\hat{q}$, and this process is recursively applied to find multiple change points. This class of methods is broadly referred to as ``top down" \citep{keogh2004segmenting} or ``forward selection" \citep{niu2016multiple} algorithms. Typical adaptations of this method include the addition of a minimum sample size or distance between change points $\delta$, and hypothesis testing to evaluate the candidate change point(s). However, binary segmentation has some issues. First, it is inefficient and equivalent to carrying out a brute-force search of change points which becomes exacerbated when each model fitment itself is expensive. Second, it has been shown that it is not effective for the multiple change point setting \citep{fryzlewicz2014wild, niu2016multiple}.

Instead of evaluating all possible time points, we propose a novel efficient method to identify multiple candidate change points which considers a binary search method that has been adapted for change point detection. Binary search, also called half-interval search \citep{williams}, logarithmic search \citep{knuth}, or binary chop \citep{butterfield}, is a foundational algorithm typically used to find the position of values in a sorted list. Since binary search is fast, scalable, and easy to implement, it works well for our multivariate high-dimensional time series setting and application. We provide a description of the binary search method for change point detection in Algorithm \ref{alg:binsearch}.

\begin{algorithm}[ht]
\SetAlgoLined

\DontPrintSemicolon

\KwInputs{$\boldsymbol{X}, \delta, model, L$}

Initialize the search space $Q \gets [1+\delta, ... , T-\delta]$

\While{$|Q| > 1$}{

    Define the midpoint of the time series; $t_{\text{mid}} \gets \frac{\sup Q - 1}{2}$
    
    Define the left and right time series segments, $\boldsymbol{V}^{L} \gets \boldsymbol{X}_{\inf Q:t_\text{mid},}$; $\boldsymbol{V}^{R} \gets \boldsymbol{X}_{t_\text{mid}:\sup Q,}$, respectively
    
    Fit and evaluate model on segments, $l^{L} \gets L_{model}(\boldsymbol{V}^{L})$ and $l^{R} \gets L_{model}(\boldsymbol{V}^{R})$
    
    Narrow the search space $Q$ into the time indices associated with the $\boldsymbol{V}$ of greater loss, $l$
}

Define the candidate change point, $\hat{q} \gets Q$
\label{alg:binsearch}

\Return $\hat{q}$
\caption{Binary search algorithm (\texttt{binsearch}) adapted for change point detection for input time series $\boldsymbol{X} \in \mathbb{R}^{T \times p}$. Any $model$ and loss metric, $L$, of choice can be used.}
\end{algorithm}

\begin{figure}[!ht]
\begin{center}
  \includegraphics[width=1\linewidth]{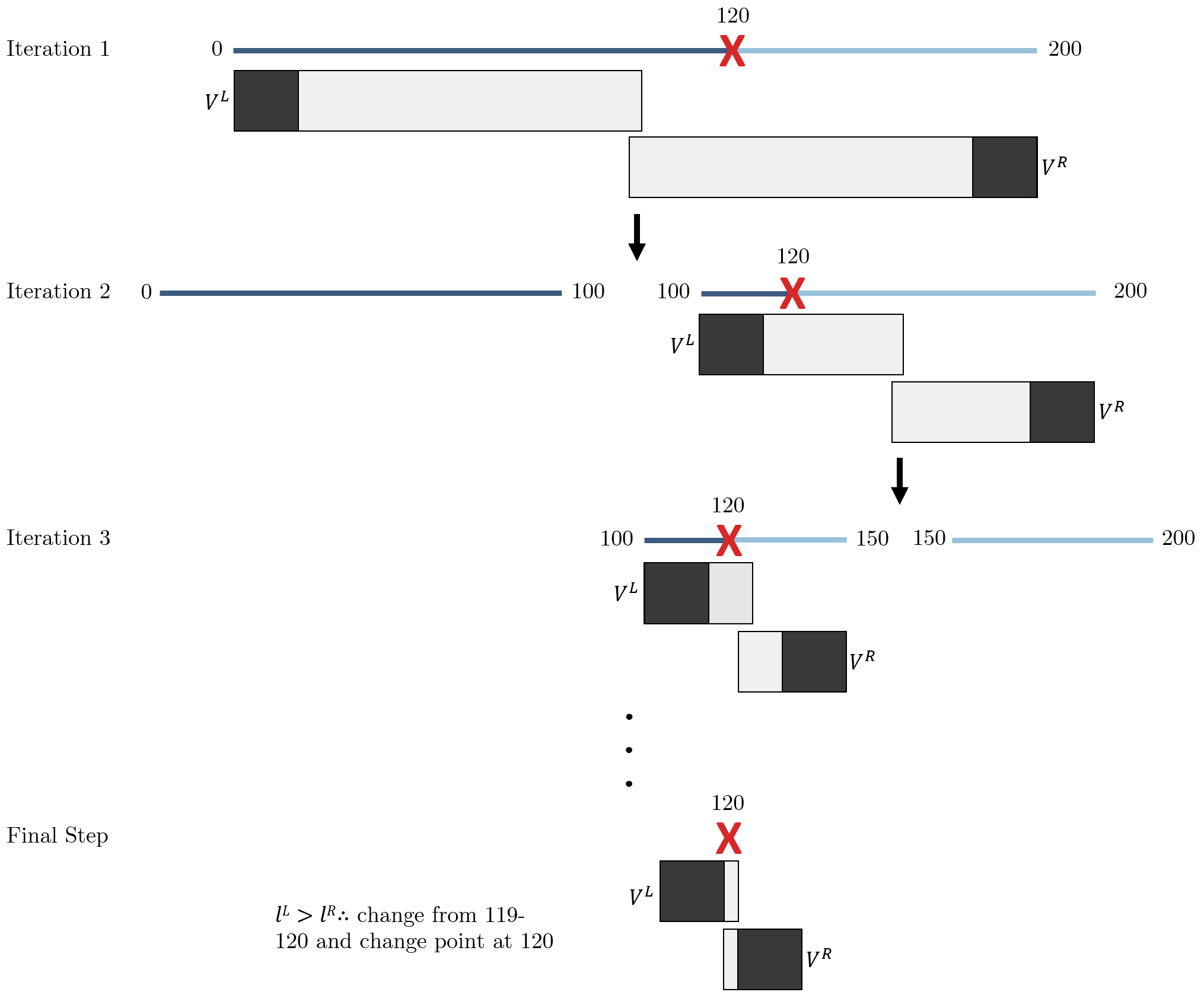}
  \caption{A graphic of the modified binary search procedure for FaBiSearch. The multivariate time series ($T = 200$) denoted by the dark blue and light blue lines, is progressively cut in half to find the true change point ($\tau = 120$) denoted by the red cross. Below each time series are the two (left and right) blocks of data where NMF is estimated.  Each block consists of $\delta$ worth of padding at the outermost edges (black), and the overlap of the two blocks (grey) in the middle.}
  \label{fig:binsearch}
\end{center}
\end{figure}

To describe our binary search method, let us first consider a one change point setting. Implicit in binary segmentation is the idea that for some $t \in \tau$ and some shift away from this ($\varepsilon$), $L_{model}(\boldsymbol{X}_{1:t+\varepsilon,}) + L_{model}(\boldsymbol{X}_{t+\varepsilon+1:T,}) > L_{model}(\boldsymbol{X}_{1:t,}) + L_{model}(\boldsymbol{X}_{t+1:T,})$. More generally, the sum of losses of two segments increases as the distance from the true change point increases. This can be attributed to an inferior model fit and therefore higher loss in the segment which contains the candidate change point, which is loosely related to the idea of optimistic search in \cite{kovacs2020optimistic}. This is precisely what underlies the mechanisms of Algorithm \ref{alg:binsearch}, wherein at each iteration, we consider two segments $\boldsymbol{V}^L$ and $\boldsymbol{V}^R$ which overlap at the midpoint of $\boldsymbol{X}$, and narrow the search space to the segment with higher loss since it is more likely to contain a candidate change point. Each iteration of this process cuts the search space $Q$ approximately in half until the length of the time index equals $1$, which signals the end of the process. The candidate change point, $\hat{q}$, is simply the remaining time index in $Q$. This is summarized with an example graphic in Figure \ref{fig:binsearch}. Similar to binary segmentation, once $\hat{q}$ is detected, $\boldsymbol{X}$ is then split into two child segments $\boldsymbol{X}_{1:\hat{q}}$ and $\boldsymbol{X}_{\hat{q}+1:T}$. Naturally then, the algorithm can be applied recursively to each child segment to find multiple change points, as long as $|T| \geq 1$ in the child segments.

The advantages of this approach are two-fold. First, by iteratively halving the search space, binary search improves upon the sequential, exhaustive search of binary segmentation by reducing the worst case search size from $O(T)$ to $O(\log{T})$. This approach allows for change point detection in larger data sets more computationally feasible. Second, by successively narrowing the search space at each iteration, the influence of other change points in a multiple change point context becomes diminished. As a result, this approach also improves upon the accuracy of change point detection. For finite sample performance we refer readers to \cite{ondrus2021} which provides a comparison of binary search to binary segmentation in a multiple change point simulation study.

\subsection{Refitting segments and statistical inference} \label{subsec:inf}
Once the binary search algorithm has been exhausted, and we have detected candidate change points $\hat{Q}$, we carry out a refitting procedure. Since our strategy is to overestimate the number of change points by recursively segmenting $\boldsymbol{X}$, we next prune the number of candidate change points detected using the following method which fits more broadly within the class of permutation tests \citep{fisher1936design, antoch2001permutation}. This two-phase approach is closely related to ``tree pruning" in the context of classification and regression trees \citep{breiman2017classification}. For the general case, the $m$ candidate change points arranged in ascending order are defined as $\hat{Q} = \{\hat{q}_{1}, \hat{q}_{2},...,\hat{q}_{m}\}$. Then, we divide $\boldsymbol{X}$ into the set of $m+1$ predicted stationary segments $\hat{S} = \{\boldsymbol{\hat{S}_1}, \boldsymbol{\hat{S}_2}, ..., \boldsymbol{\hat{S}_{m+1}}\}$, where the $i$th segment is defined as $\hat{\boldsymbol{S_i}} = \boldsymbol{X}_{b_i-1:b_i,}$ where $b = \{1, \hat{Q}, T\}$ is the set of candidate change points inclusive of the end points $1$ and $T$.

For each $\hat{q}_i$, we define the pair $\boldsymbol{\hat{S}_i}$ and $\boldsymbol{\hat{S}_{i+1}}$ which are the stationary segments immediately before and after the $i$th candidate change point. Next, we define $\boldsymbol{\hat{S'}_{i}}$ and $\boldsymbol{\hat{S'}_{i}}$ which are obtained from permuting $\boldsymbol{\hat{S}_{i}} \cup \boldsymbol{\hat{S}_{i+1}}$ over time points, and splitting at $\hat{q}_i$. This step effectively disrupts any existing piece-wise temporal ordering and provides a reference of no change point structure. The loss at $\hat{q}_i$ is $L_{model}(\boldsymbol{\hat{S}_{i}}) + L_{model}(\boldsymbol{\hat{S}_{i+1}})$, while the permuted loss is $L_{model}(\boldsymbol{\hat{S'}_{i}}) + L_{model}(\boldsymbol{\hat{S'}_{i+1}})$. We repeat this procedure for $n_{reps}$ number of repetitions, generating samples of the the unpermuted and permuted loss, $l_i$ and $l'_i$, respectively.

To determine whether a candidate change point is a change point, we are interested in the following hypothesis:

\[
\begin{array}{c}
H_{0} : \mu(l_{i}) \geq \mu(l'_{i}) \\
H_{a} : \mu(l_{i}) < \mu(l'_{i}).
\end{array}
\]
\noindent We use Welch's $t$-test and construct the test statistic as follows:

\[
t = \frac{\bar{l}_{i} - \bar{l'}_{i}}{\sqrt{\frac{s_{i}^{2}}{N_{i}} + \frac{s^{'2}_{i}}{N^{'}_{i}}}}.
\]

\noindent In the case we have multiple candidate change points, we adjust the $p$-value in each statistical test for multiple comparisons using the method from \cite{benjamini}. Each candidate change point in $\hat{q}$ becomes a change point $\hat{q}^*$ if and only if we reject $H_0$ for a given $\alpha$. This process is repeated for each candidate change point in $\hat{Q}$. The complete FaBiSearch method using NMF is described in Algorithm \ref{algo:FaBiSearch}. 

\begin{algorithm}
\caption{FaBiSearch algorithm for multiple change point detection. }

\DontPrintSemicolon
\SetAlgoLined

\KwInputs{$\boldsymbol{X}$, $\alpha$, $\delta$, $n_{run}$, $n_{reps}$, $r$}

	\label{algo:FaBiSearch}
	
\uIf{$r$ is not given}
	{
	\nl find $r^*$ using \texttt{optrank}($\boldsymbol{X}$) for subsequent uses of \texttt{NMF}
	}

\nl apply \texttt{binsearch}($\boldsymbol{X}, \delta$, \texttt{NMF}, KLD) recursively to find candidate change points $\hat{Q} = (\hat{q}_{1}, \hat{q}_{2},...,\hat{q}_{m})$, define $b = \{1, \hat{Q}, T\}$

\For{$i \gets 1$ \KwTo $m$}
    {
    \nl obtain $\boldsymbol{\hat{S'}_{i}}$ and $\boldsymbol{\hat{S'}_{i+1}}$ from $\boldsymbol{\hat{S}_{i}} \cup \boldsymbol{\hat{S}_{i+1}}$
    
    \For{$j \gets 1$ \KwTo $n_{reps}$}
	{
	\nl calculate $l_{i}(j)$ and $l'_{i}(j)$
	}

    apply Welch's $t$-test; \uIf{$\mu(l_{i}) < \mu(l^{'}_{i})$ at given $\alpha$}
	{
	\nl $\hat{q}^*_{i} \gets \hat{q}_{i}$
	}
   
   }

\Return{$\hat{Q}^* = \{\hat{q_1}^*,..., \hat{q_s}^*\}$}
   
\end{algorithm}

\section{Estimating stationary networks}
\label{sec:netest}

To estimate the stationary networks between change points, we introduce an NMF-based method for computing an adjacency matrix. For each stationary block of data, $\boldsymbol{S_i}$, the first step is to estimate NMF. Cluster membership is determined by the $r\times p$ coefficient matrix, $\boldsymbol{H}$, wherein a particular column of $\boldsymbol{H}$ (or a time series) is assigned to the cluster which has the highest coefficient value in the rows of $\boldsymbol{H}$. For each run in $n_{runs}$, an adjacency matrix,
\begin{equation*}
\boldsymbol{A}_{ij} = \begin{cases}
1, & \mbox{if } i, j \mbox{ are in the same cluster};\\
0, & \mbox{otherwise}
 \end{cases}
\end{equation*}
\noindent is generated representing the cluster membership of each node (which represents each time series). By taking the average of all adjacency matrices generated over $n_{runs}$, we obtain the corresponding consensus matrix,
\[
\boldsymbol{C} = \mu(\boldsymbol{A}_1 \text{ ,..., } \boldsymbol{A}_{n_{runs}}), ~~ 0 \leq \boldsymbol{C}_{ij} \leq 1 ~~~\forall i,j \text{.}
\]

\noindent This procedure has the advantage that it can combine results across $n_{run}$, which may be unfavourable on their own, into an overall matrix where each entry denotes the probability of two nodes being clustered together. This is similar to stability selection in \cite{meinshausen} and bootstrapping in \cite{zhu2018}.  From here, we can apply a clustering algorithm (such as hierarchical clustering with complete-linkage) to classify cluster membership amongst nodes in the consensus matrix. This provides a clustering tree which we then cut at a predetermined number of clusters. The final product is an adjacency matrix computed as an ensemble of multiple NMF runs. The new method is shown in Figure \ref{fig:graphingmethod}. While this method provides an intuitive way to organize nodes into clusters, it requires a prespecified number of communities to be defined. In some cases or applications, this may not be known and it may not be a convenient way of interpreting interactions amongst nodes. Further, there is no way to adjust for a level of sparsity and thus many networks may be too dense to easily interpret. As such, we can also define relationships amongst nodes using the consensus matrix values and using a prespecified threshold, $\lambda$, to control the sparsity/density of the resulting adjacency matrix:

\begin{figure}[ht]
\begin{center}
  \includegraphics[width=1\linewidth]{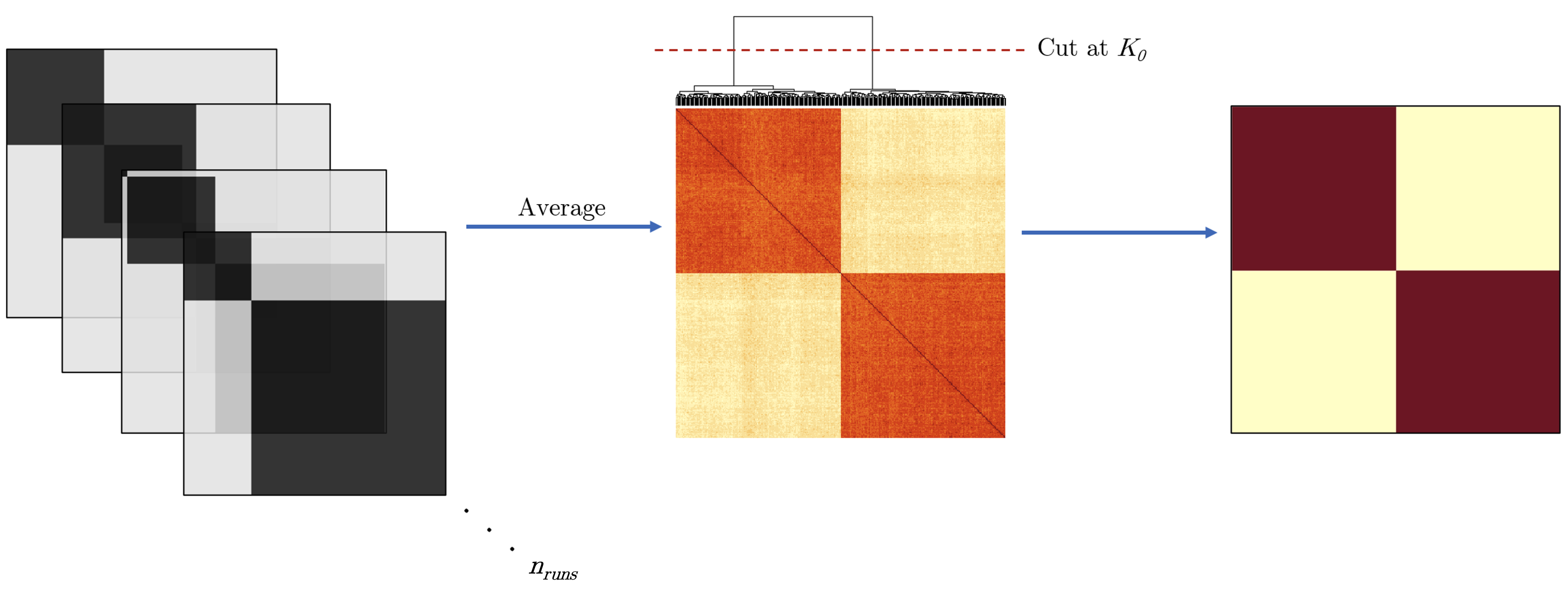}
  \caption{The workflow for visualizing networks between pairs of change points (stationary blocks). Individual adjacency matrices are calculated over $n_{runs}$ and then averaged to find the consensus matrix. Using a clustering algorithm the consensus matrix is determined. Finally, the tree is cut at a prespecified $K_{0}$ value and the nodes are then assigned cluster membership.}
  \label{fig:graphingmethod}
\end{center}
\end{figure}

\begin{equation*}
\boldsymbol{A}_{ij} = \begin{cases}
1, & \mbox{if } \boldsymbol{C}_{ij} > \lambda;\\
0, & \mbox{otherwise}.
 \end{cases}
\end{equation*}

\section{Data}\label{sec:data}

\subsection{Simulation study setup} \label{sec:sim_setup}
We apply our FaBiSearch method to multiple simulations across various dimensions.  As a measure of accuracy of the detected locations in time, we compare the  the detected change points to the location of the true change points using the scaled Hausdorff distance, $$d_H = n_s^{-1}\max\left\lbrace \max_j\min_k\left|q_j-\hat{q}_k\right|,\max_k\min_j\left|q_j-\hat{q}_k\right|\right\rbrace,$$ where $n_s$ is the length of the largest segment, $\hat{q}_k$ are the estimated change points and $q_j$ are the true change points.  The objective is to find a model that minimizes the scaled Hausdorff distance. 

For each simulation, we generated 100 iterations. For FabiSearch: we used the \texttt{R} package \texttt{NMF} \citep{Gaujoux2010} to implement NMF, the generalized Kullback-Leibler Divergence (\ref{eqn:KD}) for the loss measure, a minimum distance between change points of $\delta = 35$, the number of runs $n_{run} = 50$, the number of permutations $n_{reps} = 100$ and a significance level of $\alpha = 0.01$ for the inference step in Section~\ref{subsec:inf}. For the simulations, we assume the number of clusters (or latent factors) is unknown, hence, we did not pre-specify rank and thus let FaBiSearch find the optimal value for the rank, $r$.  

We compare our FaBiSearch method to the Network Change Point Detection (NCPD) method \citep{Cribben2017} as it is the only competing high-dimensional method that has software available (which is not the case for the majority of the methods mentioned in the introduction). NCPD finds change points in the network (or community) structure of multivariate high-dimensional time series data. Unlike FaBiSearch, NCPD uses binary segmentation and applies spectral clustering to find the communities based on the computed correlation matrix. The difference between the communities before and after a candidate change point are calculated using the principal angle and a stationary bootstrap is used for inference.   In the simulations, for NCPD, we used a pre-specified number of clusters, $K$, of one greater than the actual ($K_{0} + 1$) as suggested in the paper, 1000 stationary bootstrap replications, a minimum distance between change points of $\delta = 50$, and a significance level of $\alpha = 0.05$.  These are the default settings of the method.

\subsection{Simulated data} \label{sec:sims}

\begin{figure}[!ht]
\begin{center}
  \includegraphics[width=0.7\linewidth]{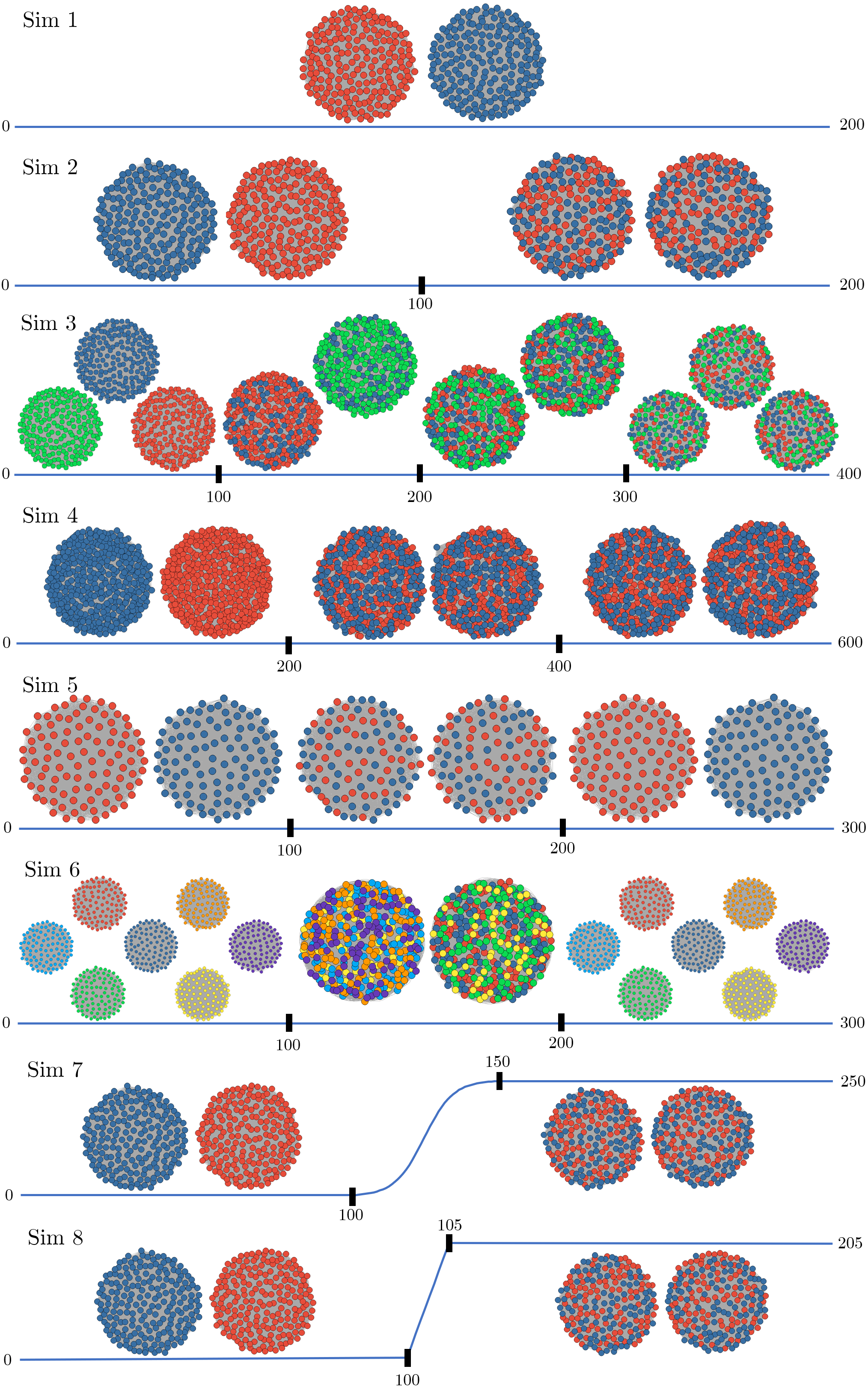}
  \caption{The true network structure of the simulations. The time series are represented by nodes.  The colors of nodes indicate how the cluster membership evolves throughout the time series. Solid black vertical lines on the time axes indicate the location of the true change points. For Simulations 7 and 8, the slowly transitioning simulations, the shape of the change segments (100:150 and 100:105, respectively) denote the weighting (sigmoidal and linear, respectively) of the clustering structure before and after the segment.}
  \label{fig:sims}
\end{center}
\end{figure}

We now describe the simulated data. We chose simulations to emulate qualities of fMRI time series data.  While the data are simulated from various models, we display the structure between the time series (or nodes) using graphs (or networks, which we use interchangeably) in Figure \ref{fig:sims}.   A graph  consists of a set of nodes $N$ and corresponding edges $E$ that connect pairs of nodes.  Here, each node represents a time series, or ROI, and edges encode dependencies. In the fMRI setting, a missing edge indicates a lack of functional connectivity between corresponding regions.  In Figure \ref{fig:sims}, the time series are represented by nodes and the colors of the nodes indicate  cluster (or community) membership. 

We progressively increased the difficulty and complexity of the simulations in order to characterize the power of FaBiSearch. All simulations are multivariate high-dimensional time series with a defined clustering (or community) structure, where $K_{0}$ is the true number of clusters/communities. 

 \noindent \textbf{Simulation 1: }  There are no change points. $T = 200$, $p = 400$ time series, and the number of clusters is $K_{0} = 2$.  The data are generated from the multivariate Gaussian distribution $\mathcal{N}(0, \boldsymbol{\Sigma})$, where 
\begin{equation*}
\boldsymbol{\Sigma}_{ij} = \begin{cases}
0.75, & \mbox{if } i\neq j \mbox{ and } i, j \mbox{ are in the same cluster};\\
 1, & \mbox{if } i = j;\\
0.20, & \mbox{otherwise}.
 \end{cases}
\end{equation*}
This simulation is similar to a steady state fMRI time series, where the network structure does not change over time.

\noindent \textbf{Simulation 2: }  We consider one change point. $T = 200$, $p = 400$, and the number of clusters is $K_{0} = 2$. The change point occurs at $\tau = 100$.  Data before the change point are generated from the multivariate Gaussian distribution $\mathcal{N}(0, \boldsymbol{\Sigma})$, where
\begin{equation*}
\boldsymbol{\Sigma}_{ij} = \begin{cases}
0.75, & \mbox{if } i\neq j \mbox{ and } i, j \mbox{ are in the same cluster};\\
 1, & \mbox{if } i = j;\\
0.20, & \mbox{otherwise}.
 \end{cases}
\end{equation*}
Data after the change point are generated from the same multivariate Gaussian distribution $\mathcal{N}(0, \boldsymbol{\Sigma})$ but the node labels are randomly reshuffled.

\noindent \textbf{Simulation 3: } We consider three change points.  $T = 400$, $p = 600$ time series.  The change points occur at $\tau = \{100, 200, 300\}$.  Data are generated from the multivariate Gaussian distribution $\mathcal{N}(0, \boldsymbol{\Sigma})$, where
\begin{equation*}
\boldsymbol{\Sigma}_{ij} = \begin{cases}
0.75, & \mbox{if } i\neq j \mbox{ and } i, j \mbox{ are in the same cluster};\\
 1, & \mbox{if } i = j;\\
 0.20^{|i-j|}, & \mbox{if } i, j \mbox{ are not in the same cluster}.\\ 
 \end{cases}
\end{equation*}
In the first time segment, the true number of clusters $K_{0} = 3$, one of which is equally merged into the other two clusters at the first change point. Vertex labels are randomly shuffled at the second change point, while keeping $K_{0} = 2$. The true number of clusters $K_{0}$ returns to 3, by moving one third of each cluster into a new, third cluster.

\noindent \textbf{Simulation 4: } We consider 2 change points.  $T = 600$, $p = 800$ time series. The change points occur at $\tau = \{200, 400\}$.  The number of clusters is $K_{0} = 2$ throughout.  Data are generated from the multivariate Gaussian distribution $\mathcal{N}(0, \boldsymbol{\Sigma})$, where
\begin{equation*}
\boldsymbol{\Sigma}_{ij} = \begin{cases}
0.75, & \mbox{if } i\neq j \mbox{ and } i, j \mbox{ are in the same cluster};\\
 1, & \mbox{if } i = j;\\
 0.20^{|i-j|}, & \mbox{if } i, j \mbox{ are not in the same cluster}.\\ 
 \end{cases}
\end{equation*}
At both change points, half of the vertices in each cluster are chosen at random and moved to the other cluster.

\noindent \textbf{Simulation 5: } We consider 2 change points.  $T = 300$, $p = 200$ time series.  The change points occur at $\tau = \{100, 200\}$. The number of clusters is $K_{0} = 2$ throughout. Data are generated from the multivariate Gaussian distribution $\mathcal{N}(0, \boldsymbol{\Sigma})$, where 
\begin{equation*}
\boldsymbol{\Sigma}_{ij} = \begin{cases}
0.75, & \mbox{if } i\neq j \mbox{ and } i, j \mbox{ are in the same cluster};\\
 1, & \mbox{if } i = j;\\
 0.20^{|i-j|}, & \mbox{if } i, j \mbox{ are not in the same cluster}.\\ 
 \end{cases}
\end{equation*}
In the second time segment, the vertex labels are randomly shuffled from the first time segment. In the third time segment, the vertex labels are the same as the first time segment.  In this simulation, we are mimicking the ABA structure, where the subject alternates between two states.

\noindent \textbf{Simulation 6: } We consider two change points. $T = 300$, $p = 200$ time series. The change points occur at $\tau = \{100, 200\}$. Data are generated from the multivariate Gaussian distribution $\mathcal{N}(0, \boldsymbol{\Sigma})$, where
\begin{equation*}
\boldsymbol{\Sigma}_{ij} = \begin{cases}
0.75, & \mbox{if } i\neq j \mbox{ and } i, j \mbox{ are in the same cluster};\\
 1, & \mbox{if } i = j;\\
 0.20^{|i-j|}, & \mbox{if } i, j \mbox{ are not in the same cluster}.\\ 
 \end{cases}
\end{equation*}
In the first and third time segments, $K_{0} = 7$ and vertex labels are the same. In the second time segment, the vertex labels from clusters 1 to 3 and 5 to 7 in the first time segment are grouped into two larger clusters. Half of vertex labels from cluster 4 in the first time segment are put into each of these two larger clusters.  In this simulation, we are mimicking the ABA structure, where the subject alternates between two states.  The number of clusters match the real fMRI data considered in Section~\ref{subsec:harry_res}.

\noindent \textbf{Simulation 7: }  We consider one change point.  $T = 200$, $p = 400$ time series.  For the first stationary segment, ($1:100$), data are generated from the multivariate Gaussian distribution $\mathcal{N}(0, \boldsymbol{\Sigma})$, where
\begin{equation*}
\boldsymbol{\Sigma}_{ij} = \begin{cases}
0.75, & \mbox{if } i\neq j \mbox{ and } i, j \mbox{ are in the same cluster};\\
 1, & \mbox{if } i = j;\\
0.20, & \mbox{otherwise}.
 \end{cases}
\end{equation*}
In the second stationary segment ($150:250$) vertices labels are randomly reshuffled.  The change segment occurs at $\tau = 100:150$, which is a weighted combination of stationary segment 1 and 2. Weighting is defined by $w_{c} = \frac{1}{1+e^{-0.2*(t-25)}}$, where $t$ is the number of time points after $100$. $K_{0} = 2$.  In this simulation, we are mimicking a very slow transition between states.

\noindent \textbf{Simulation 8: }  We consider one change point.  $T = 200$, $p = 400$ time series.  $K_{0} = 2$.  For the first stationary segment, ($1:100$), data are generated from the multivariate Gaussian distribution $\mathcal{N}(0, \boldsymbol{\Sigma})$, where
\begin{equation*}
\boldsymbol{\Sigma}_{ij} = \begin{cases}
0.75, & \mbox{if } i\neq j \mbox{ and } i, j \mbox{ are in the same cluster};\\
 1, & \mbox{if } i = j;\\
0.20, & \mbox{otherwise}.
 \end{cases}
\end{equation*} 
In the second stationary segment ($105:205$) vertices labels are randomly reshuffled. The change segment occurs at $\tau = 100:105$, which is a weighted combination of stationary segment 1 and 2. Weighting is defined by $w_{c} = \frac{t}{10}$, where $t$ is the number of time points after $100$.  In this simulation, we are mimicking a transition between states. 

Simulations 2, 3, and 4, were taken directly from \cite{Cribben2017}, for comparison purposes.

\subsection{fMRI study setup}
We also applied FaBiSearch to two fMRI data sets: a resting-state fMRI data set and a task-based fMRI data set. NMF, by definition, requires the input matrix to be non-negative however fMRI data has no such restriction. To circumvent this issue, we shift the data to make it positive by adding the same positive value to all entries of the input matrix. This ensures that the input matrix $\boldsymbol{X}$ contains only positive values while also preserving individual variability and the covariance between ROIs.  We used the same inputs for FaBiSearch for analyzing the fMRI data set as in the simulation study (end of Section \ref{sec:sim_setup}).

\subsection{Resting-state fMRI data} \label{section:rs_fmridata}
This data set includes 25 participants (mean age of 29.44 $\pm$ 8.64, 10 males and 15 females) scanned at New York University over three visits (\url{http://www.nitrc.org/projects/nyu_trt}). For each visit, participants were asked to relax, remain still, and keep their eyes open. A Siemens Allegra 3.0-Tesla scanner was used to obtain the resting-state scans for each participant, however we considered only the second and third visits because they were less than an hour apart. Each visit consisted of 197 contiguous EPI functional volume scans with time repetition (TR) of 2000ms, time echo (TE) of 25ms, flip angle (FA) of 90$^{\circ}$, 39 number of slices, matrix of $64\times64$, field of view (FOV) of 192mm, and voxel size of $3\times3\times3$mm$^3$. Software packages \texttt{AFNI }(\url{http://afni.nimh.nih.gov/afni}) and \texttt{FSL }(\url{http://www.fmrib.ox.ac.uk}) were used for preprocessing. Motion was corrected using \texttt{FSL}'s \texttt{mcflirt} (rigid body transform, cost function normalized correlation, and reference volume the middle volume). Normalization into the Montreal Neurological Institute (MNI) space was performed using \texttt{FSL}'s \texttt{flirt} (affine transform, cost function, mutual information). Probabilistic segmentation was conducted to determine white matter and cerebrospinal fluid (CSF) probabilistic maps and was obtained using \texttt{FSL}'s \texttt{fast} with a threshold of 0.99. Nuisance signals (the six motion parameters, white matter signals, CSF signals, and global signals) were removed using \texttt{AFNI}'s \texttt{3dDetrend}. Volumes were spatially smoothed using a Gaussian kernel and FWHM of 6mm with \texttt{FSL}'s \texttt{fslmaths}. We used the work of \cite{Gordon2016} to determine the ROI atlas. The cortical surface is parcellated into 333 areas of homogenous connectivity patterns, and the time course for each is determined by averaging the voxels within each region for each subject. Regional time courses were then detrended and standardized to unit variance. Lastly, a fourth-order Buttterworth filter with a 0.01-0.10 Hertz pass band was applied.

\subsection{Task-based fMRI data} \label{section:task_fmridata}
This data set includes 8 participants (ages 18-40) scanned at the Scientific Imaging and Brain Imaging Center at Carnegie Mellon University (\cite{wehbe2014}: \url{http://www.cs.cmu.edu/~fmri/plosone/}). Subjects in the study were asked to read Chapter 9 of \textit{Harry Potter and the Sorcerer’s Stone} \citep{Rowling2012}. All subjects had previously read the book or seen the movie. The words of the story were presented in rapid succession, where each word was presented one by one at the center of the screen for 0.5 seconds in black font on a gray background. A Siemens Verio 3.0T scanner was used to acquire the scans, utilizing a T2* sensitive echo planar imaging pulse sequence with repetition time (TR) of 2s, time echo (TE) of 29 ms, flip angle (FA) of 79$^{\circ}$, 36 number of slices and $3\times3\times3$mm$^3$ voxels. Data was preprocessed as described in \cite{wehbe2014,xiong}. For each subject, functional data underwent realignment, slice timing correction, and co-registration with the subject’s anatomical scan, which was segmented into grey and white matter and cerebrospinal fluid. The subject’s scans were then normalized to the MNI space and smoothed with a 6 $\times$ 6 $\times$ 6mm Gaussian kernel smoother. Data was then detrended by running a high-pass filter with a cut-off frequency of 0.005Hz after being masked by the segmented anatomical mask. Finally, the Gordon brain atlas \citep{Gordon2016} was again used to extract ROIs. The final time series for the task-based data contained 4 runs (324, 337, 264, and 365 time points) of 333 ROIs for each subject.

%=======================================================================================================
\section{Results} \label{sec:results}
\subsection{Simulation results}

\begin{figure}[ht]
\begin{center}
  \includegraphics[width=1\linewidth]{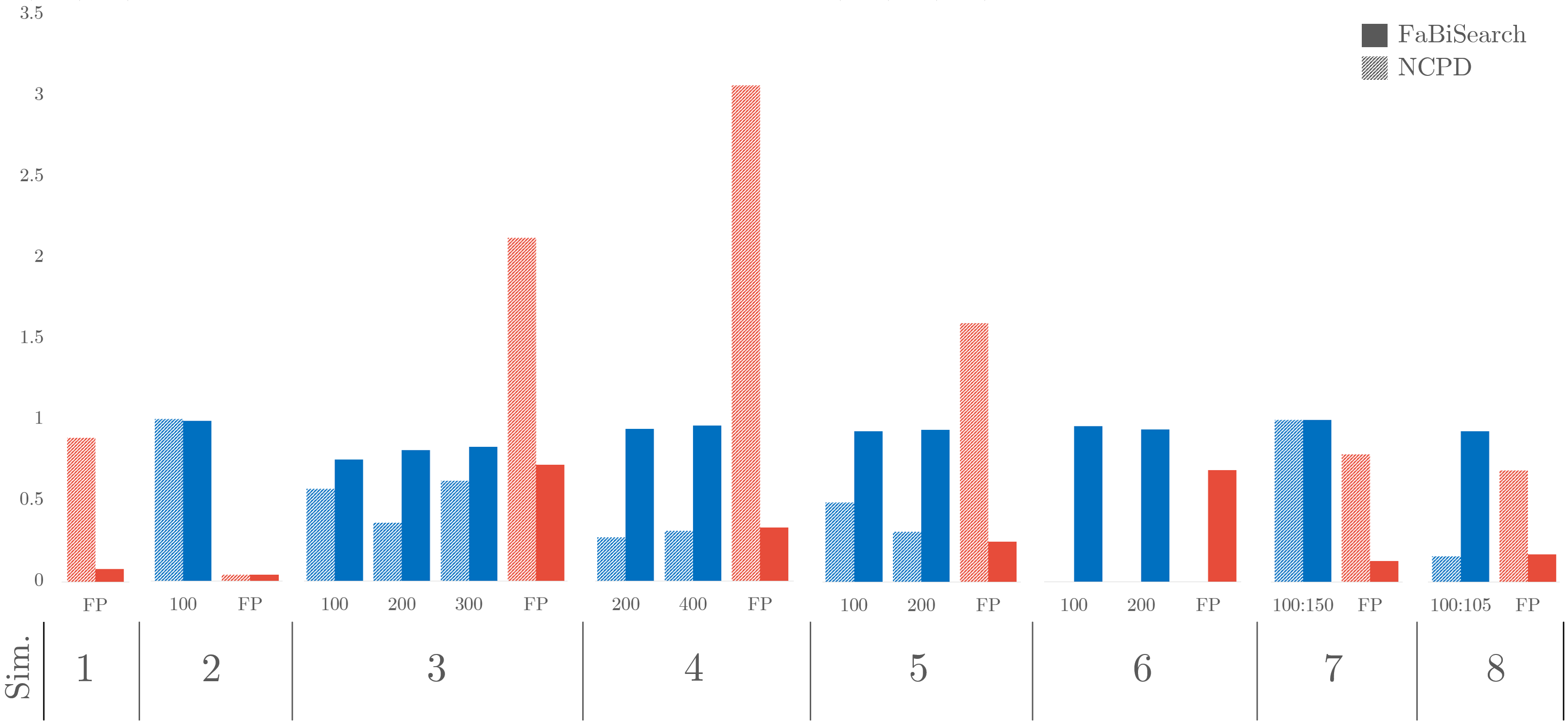}
  \caption{A plot of the true positive (blue bars) and false positive (red bars) rates for NCPD (striped fill) and FaBiSearch (solid fill) across all the simulations using the $\pm10$ window.  The $x$-axis represents the simulation number and the $y$-axis represents the TP (blue bars) and FP (red bars) rates.  The tables for both the $\pm10$ and the $\pm1$ windows can be found in the Appendix. }
  \label{fig:simresults}
\end{center}
\end{figure}

In this section, we present the simulation and fMRI results.  In both simulations and fMRI data, we assumed for FabiSearch that the number of clusters (or latent factors) are unknown and have to be estimated in each case, while for NCPD we assumed a pre-specified number of clusters, $K$, of one greater than the actual ($K_{0} + 1$), hence optimizing more in favor of NCPD.  We evaluate the change point results by calculating the true and false positives within two margins ($\pm10$ and $\pm1$ time points). A FP rate of 1 indicates that 1 FP change point is detected in all the 100 simulations, on average.  Hence, a FP rate greater than 1 is possible as more than 1 FP change point can be detected across the entire time course. 

Overall, across the simulations, FaBiSearch method outperforms the competitor NCPD \citep{Cribben2017} with the results displayed in Figure \ref{fig:simresults} (and in Table \ref{tab:simresultstable} in the Appendix).  More specifically, in Simulation 1, where there are no change points, FaBiSearch obtains a small false positive (FP) rate of 0.08, compared to NCPD's 0.89.  In Simulation 2, the two methods perform similarly on detecting one change point. FaBiSearch obtains a true positive (TP) change point rate of 0.99 across the 100 iterations while NCPD achieves a TP rate of 1, and the FP rates are the same. However, once the number of time series and the complexity increases, FaBiSearch establishes a clear advantage over NCPD. For example, in Simulation 3, FaBiSearch finds 54$\%$ more true change points than NCPD while having a reduction of 34$\%$ in the number of FPs compared to NCPD. The improvement of FaBiSearch over NCPD increases further in Simulation 4, where FaBiSearch finds 1.7 times more TP change points while reducing the number of FPs to 9$\%$ of those detected by NCPD. In Simulation 5, we tested the two methods on an ``ABA'' type structure in which the first and last clustering structures are identical. This makes the change points less discernible and thus change point detection more difficult, especially for binary segmentation methods (given the similarity between any two partitions). FaBiSearch however, performs well by detecting 93.5$\%$ of the true change points, outperforming NCPD again while detecting only a fraction of the false positives. Simulation 6 has a similar structure albeit with 7 clusters instead of 2 in the first and final stationary segments. Here, FaBiSearch outperforms NCPD again on all accounts; NCPD was not able to find any change points. FaBiSearch finds 95$\%$ of the true change points while obtaining a FP frequency of just 0.69. In Simulation 7, where there is one change segment from 100 to 150 (the network is slowly transitioning between two states), FaBiSearch and NCPD perform similarly in TP rates, obtaining frequencies of 1. FaBiSearch, however, obtains a smaller FP frequency of 0.13 compared to NCPD's 0.79. A similar outcome is seen in Simulation 8, where the change segment is linearly weighted across 5 time points (the network is slowly transitioning between two states). Here, FaBiSearch outperforms NCPD with a TP rate of 0.93 compared to 0.16, and the FP rate favours FaBiSearch (0.17) over NCPD (0.69).  Overall, FaBiSearch finds the true change points more frequently, while drastically reducing the number of FPs found, compared to NCPD. Finally, it is important to point out that the dimensionality of the problem is very large, hence making the detection points difficult as indicated by the performance of NCPD. 

Across all simulations except Simulations 1 and 7, we find that the Hausdorff distance for FaBiSearch is lower than for NCPD, (see Table \ref{tab:simresultstable} in the Appendix), indicating that the change points detected by FaBiSearch are closer to the true change points, hence, FaBiSearch outperforms NCPD on another metric.  We also assessed the efficacy of NMF in estimating networks between each pair of change points (or stationary block) in the simulations. To this end, for each iteration of each simulation, we first detected the change points and hence each stationary block using FaBiSearch.  Then for each stationary block, we estimated the network using our method described in Section \ref{section:networkestimation} where we specified the true number of clusters, $K_{0}$, as the cutoff point for the resulting tree. Since these adjacency matrices are symmetric, we quantified the difference between the true adjacency matrix and the NMF calculated adjacency matrix by calculating the percent overlap of the off-diagonal elements. The results are shown in Table \ref{tab:SIMgraphingresults} (in the Appendix). The method performs well and can recover the true network structure between change points across all simulations.

\begin{figure}[ht]
\centering
\begin{tikzpicture}[scale=1, every node/.style={scale=0.5, font=\large}]
\begin{axis}[ymajorgrids=true, ymin=0, ymax=26, xmin=40, xmax=157, axis lines=middle, minor tick num=1, ytick={1,...,25}, x label style={at={(axis description cs:0.5,-0.1)},anchor=north}, y label style={at={(axis description cs:-0.05,.5)},rotate=90,anchor=south},xlabel={\Huge Time}, ylabel={\Huge Subject number}]
\pgfplotsset{ytick style={draw=none}, width=11cm, height=7cm}
\addplot [only marks, mark size=1, color=pptblue] table [x="time", y="subject", col sep=comma] {tables/gordon2table.csv};
\addplot [only marks, mark size=1, color=pptred] table [x="time", y="subject", col sep=comma] {tables/gordon3table.csv};
\end{axis}
\end{tikzpicture}
\caption{The detected change points for each subject in the resting-state fMRI data set. Blue and red dots denote change points for the second and third scans, respectively.}
\label{fig:gordon2and3}
\end{figure}

\subsection{Resting-state fMRI results}
The objective of applying FaBiSearch to this data is to study the test-retest reliability and behavior of dynamic FC, hence we only consider the second and third scans, which were obtained less than an hour apart (Section \ref{section:rs_fmridata}). We did not consider the first scan as it was taken 5-11 months before the second resting-state scan.  Figure \ref{fig:gordon2and3} shows the detected change points for the resting-state fMRI data set using FaBiSearch.  Blue and red dots denote the change points from the second and third scans, respectively.  Although we are not certain of the number of change points for a resting-state fMRI experiment, as the subject is in an unconstrained state, we expect subjects to drift between different states (or functional modes), which is consistent with previous work \citep{delamillieure,doucet,Cribben2017,anastasiou}. We found that each subject has a unique set of change points.  Comparing the results across scanning sessions and between subjects, we see variability in the location and number of detected change points.  \cite{xu} analyzed this data to study the test-retest reliability of static FC, however, here we are considering the test-retest reliability of dynamic FC. Some subjects (e.g., subject 14) have quite different change points across the scanning sessions, while others (e.g., subjects 1, 10, 19, 20, 21, 22) have change points that are consistent across the two scans. 

\begin{figure}[ht]
\begin{center}
  \includegraphics[width=0.8\linewidth]{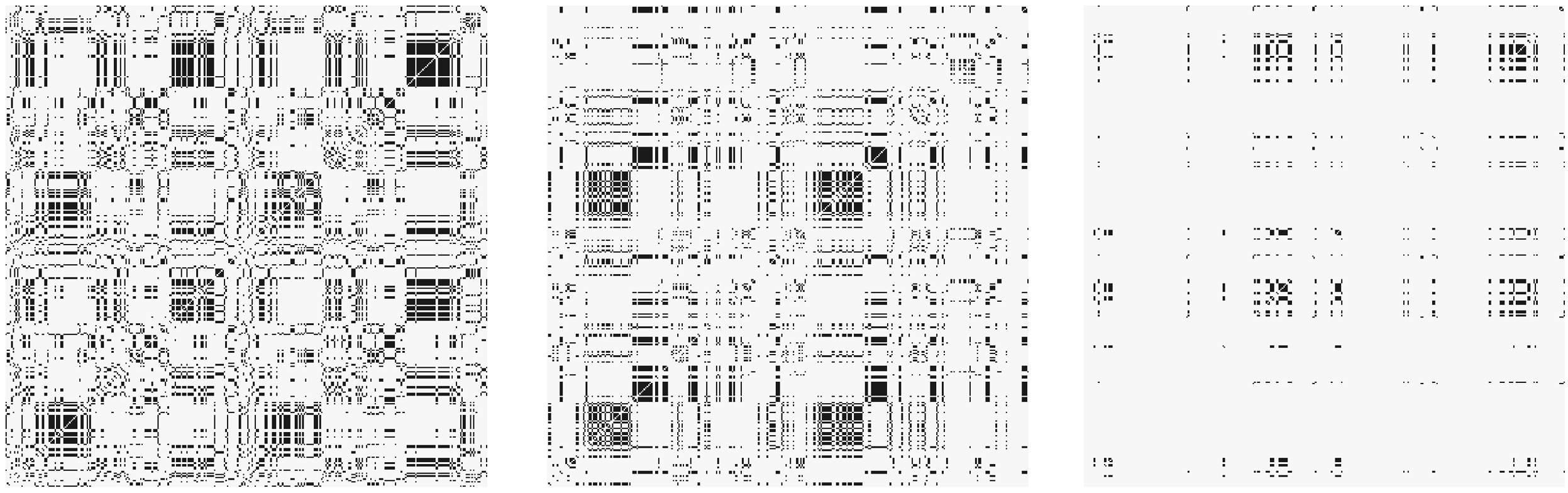}
  \includegraphics[width=0.8\linewidth]{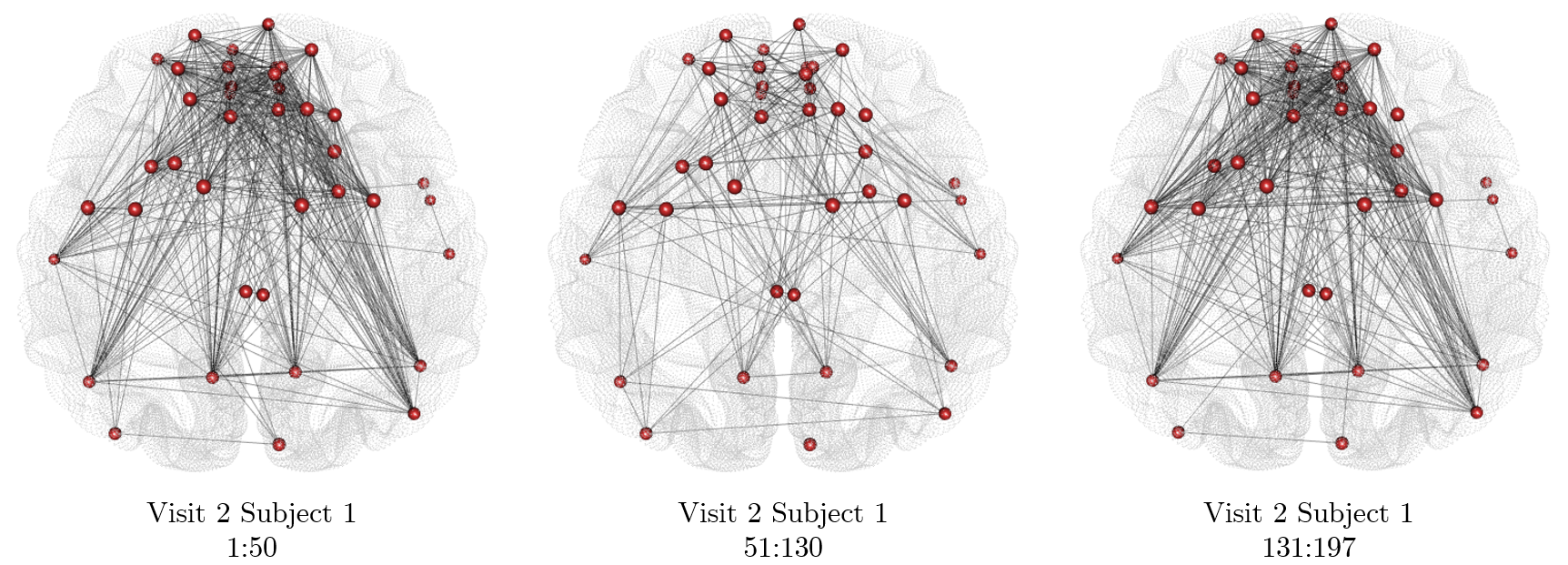}
\end{center}
  \caption{The stationary networks between each pair of change points for subject 1's second resting-state fMRI data. The top panel shows the adjacency matrices which includes all nodes, whereas the bottom panel networks are for only the ``Default" community.}
  \label{fig:rs_intrasimil}
\end{figure}

To visualize and interpret the network structure between each pair of change points detected by FaBiSearch, we estimate the networks using the NMF cluster-based method described in Section \ref{sec:netest}.  For this procedure, we fixed the number of runs ($n_{runs} = 100$) to calculate the consensus matrix.  Previous work \citep{allen} identified 7 unique resting-state networks, hence we used this prespecified cluster size.  Figure \ref{fig:rs_intrasimil} depicts the stationary networks between each pair of change points for subject 1's second fMRI resting-state scan.  The plot allows us to compare the intra-subject community structure.  There is a clear time-varying relationship between the ROIs within this network.  In fact, there are substantial changes in the clustering structure across the segments, with the first network in the top panel having a strong block diagonal structure and this progressively diminishes over the subsequent segments. Additionally, the density of the overall networks diminishes over time. In comparison however, in the bottom panels where only edges in the ``Default" community are plotted, the density appears to being high, then decrease, and then return to high again. The first and third networks in the bottom panel appear quite similarly connected (similar to a ABA type structure, see the simulations in Section \ref{sec:sim_setup}), with extremely dense and abundant connections between most of the nodes, and very few connections to the two nodes near the periphery of the rear and the three nodes on the right periphery. This is contrast to the second network in the bottom panel, where the peripheral nodes, especially the three on the right are more interconnected with the rest of the nodes.  These results provide evidence that the subject's FC, and therefore mental state, was evolving during rest. Furthermore, it also suggests that the overall networks and individual communities evolve in an independent manner.

\begin{figure}[ht]
\begin{center}
  \includegraphics[width=0.75\linewidth]{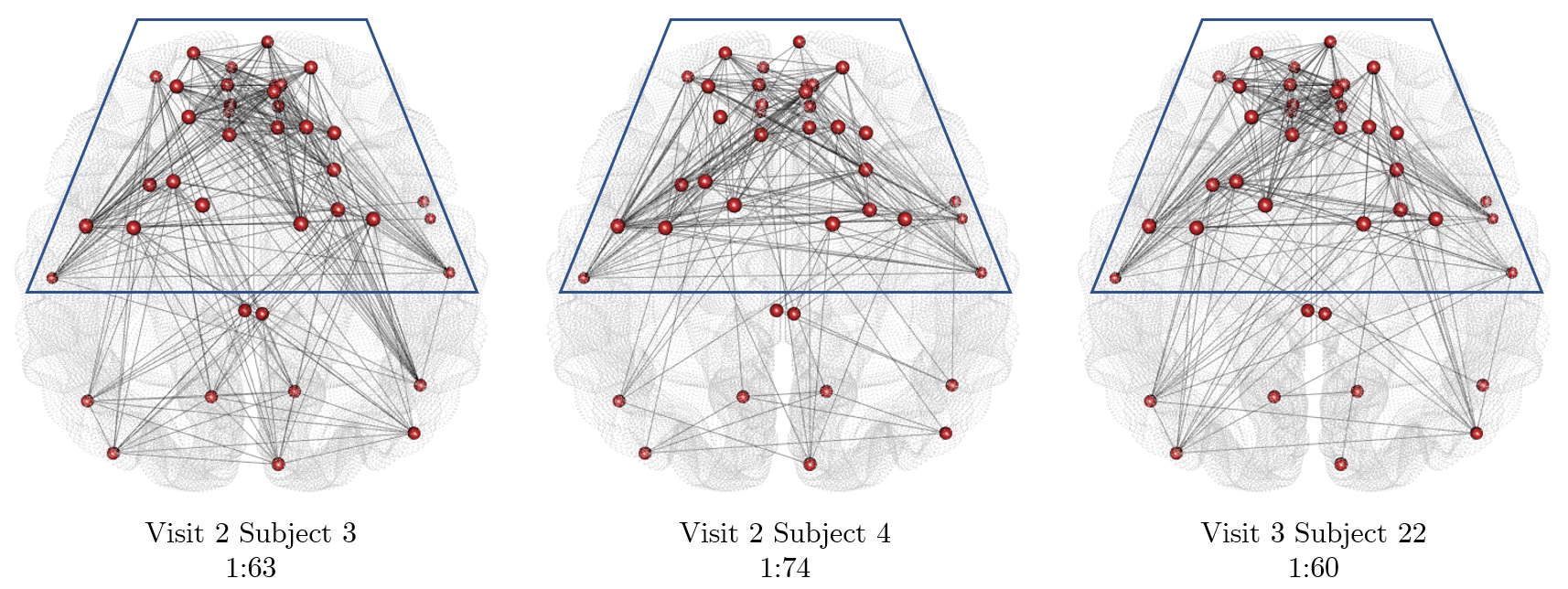}
  \includegraphics[width=1\linewidth]{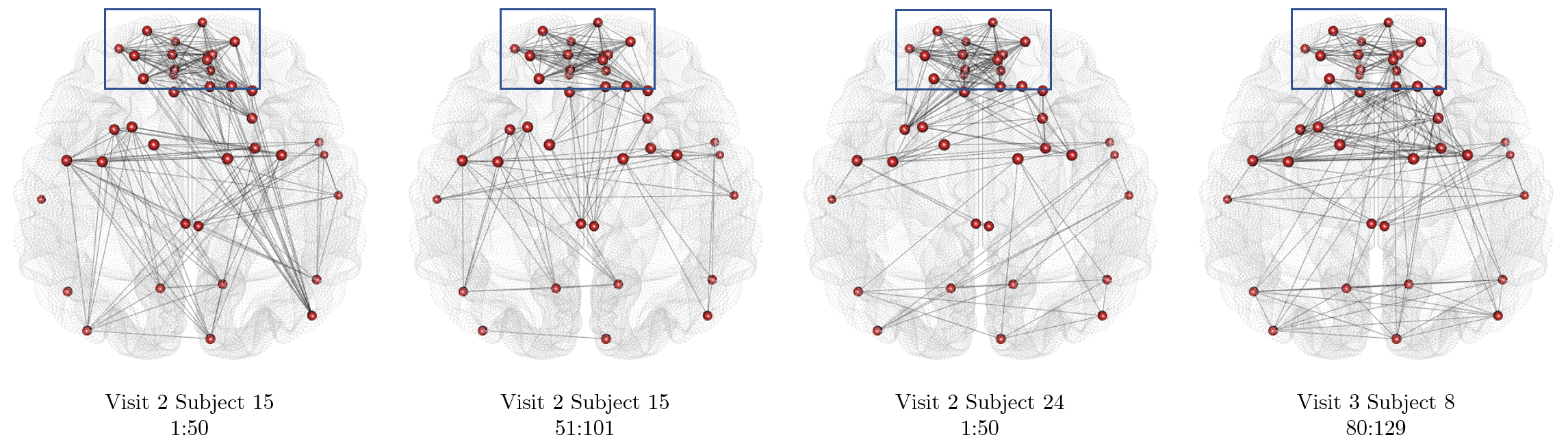}
  \caption{The estimated networks for different stationary segments across subjects and scans. Similar network motifs are outlined in blue across both top and bottom panels.}
  \label{fig:rs_intersimil}
\end{center}
\end{figure}

In Figure \ref{fig:rs_intersimil} we compare the stationary networks in the ``Default" community across both subjects and scans.  There is a clear presence of a common stable network (or subnetwork) or motifs.  This suggests that there exists certain motifs or sub-networks in time-varying FC that remain stable across subjects. Since these motifs appear across different stationary segments, it suggests that the resting-state networks do not evolve in a similar, time-dependent manner across subjects, as during a resting-state experiment, the subjects are unconstrained. These nodes appear to connect in high degree with each other. This suggests that there exist common functional states (or network features, or motifs) that are consistent across subjects in a resting-state at different time intervals which are meaningful given this is first study to consider over 300 fMRI time series ($p=333$) from a change point dynamic FC perspective.    

\subsection{Task-based fMRI results}\label{subsec:harry_res}
Figure \ref{fig:HPcpts} shows the detected change points for the task-based fMRI data set (Section \ref{section:task_fmridata}) using FaBiSearch. The results are concatenated across runs for each subject, which explains the presence of gaps, where no change points were identified. Overall, many change points were detected. In addition, there is some variability in the location and number of detected change points across subjects. Intuitively, as we considered a large number of ROIs ($p=333$) and as reading is a complex and involved task, we expect FC to be constantly evolving and different for each subject based on unique factors such as age, interpretation of the story, and familiarity with the text. There are some change points however, which are consistent across subjects, especially at critical moments and events in the storyline.

\begin{figure}[ht]
\centering
\begin{tikzpicture}[scale=1, every node/.style={scale=0.5, font=\large}]
\begin{axis}[ymajorgrids=true, ymin=0, ymax=9, xmin=30, xmax=1260, axis lines=middle, minor tick num=1, ytick={1,...,8}, x label style={at={(axis description cs:0.5,-0.1)},anchor=north}, y label style={at={(axis description cs:-0.05,.5)},rotate=90,anchor=south},xlabel={\Huge Time}, ylabel={\Huge Subject number}]
\pgfplotsset{ytick style={draw=none}, width=11cm, height=7cm}
\addplot [only marks, mark size=1, color=pptblue] table [x="time", y="subject", col sep=comma] {tables/HPcpt.csv};
\end{axis}
\end{tikzpicture}
\caption{The detected change points for each subject in the task-based fMRI data set. Blue dots denote change points, and runs 1-4 are concatenated for each subject.}
\label{fig:HPcpts}
\end{figure}

\begin{figure}[ht]
\begin{center}
  \includegraphics[width=0.9\linewidth]{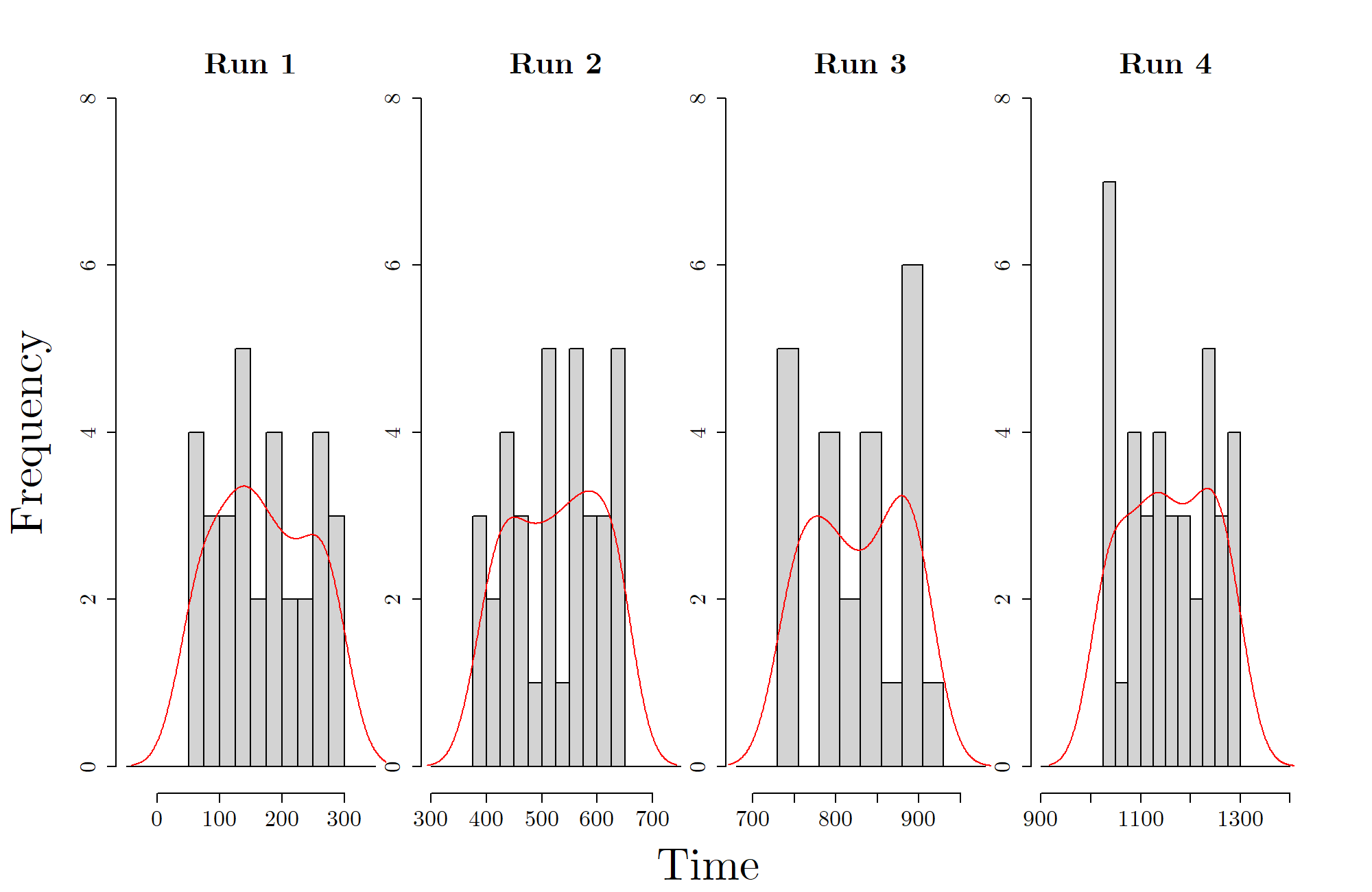}
  \caption{Histograms of the change points in the task-based fMRI study across all subjects (grey bars), overlaid with density estimates using a Gaussian kernel (red lines).}
  \label{fig:hp_cps_density}
\end{center}
\end{figure}

Figure \ref{fig:hp_cps_density} shows the histograms of the change points concatenated across subjects. Given there are only 8 subjects in this study, the histograms are very sensitive to bin size, hence we have overlaid density estimates for each of the four runs. In particular, we seek to find density estimates which show concentrated density and thus indicate a general pattern across subjects. We see that some of the smoothed peaks coincide with major events in the story. In the first run, the smoothed plot shows a peak at time point $t = 139$, which coincides with Draco taking Neville's remembrall, and precedes the first flying lesson for Harry and his classmates. The second run coincides with the following story line: after the class is left unattended by the teacher, Madam Hooch, Harry chases classmate Draco Malfoy on a broom, having never flown before. Another teacher, Professor McGonagall, spots this mischief and, instead of punishing Harry, offers him a spot as a Seeker on the Gryffindor Quidditch team. The density peak for this run is at approximately time point $t = 585$, and coincides with this offer, marking the beginning of Harry's Quidditch playing career, which is an important recurring narrative in the book series. In the third run, Harry has accepted a wizard's duel that night with arch-rival Malfoy and is on his way with Ron to meet Malfoy at the trophy room. The change points detected peaks at two time points, where the first at $t = 778$ marks a time point just after Hermoine warns Harry and Ron not to go through with the wizard's duel. The second peak, at time point $t = 879$ occurs after Hermoine surprises Harry and Ron by catching them, and scolding them for trying to break school rules. In the fourth run, the density peaks noticeably at time point $t = 1233$. In this part of the story, the three of Harry, Ron, and Hermoine are almost caught by Argus Filch, the Hogwart's caretaker, and his cat, Mrs. Norris. As the children run, they get lost and end up in a forbidden area on the third floor. The peak seen at $t = 1233$ coincides with the children running after meeting a scary and large three-headed dog.

\begin{figure}[H]
\begin{center}
  \includegraphics[width=1\linewidth]{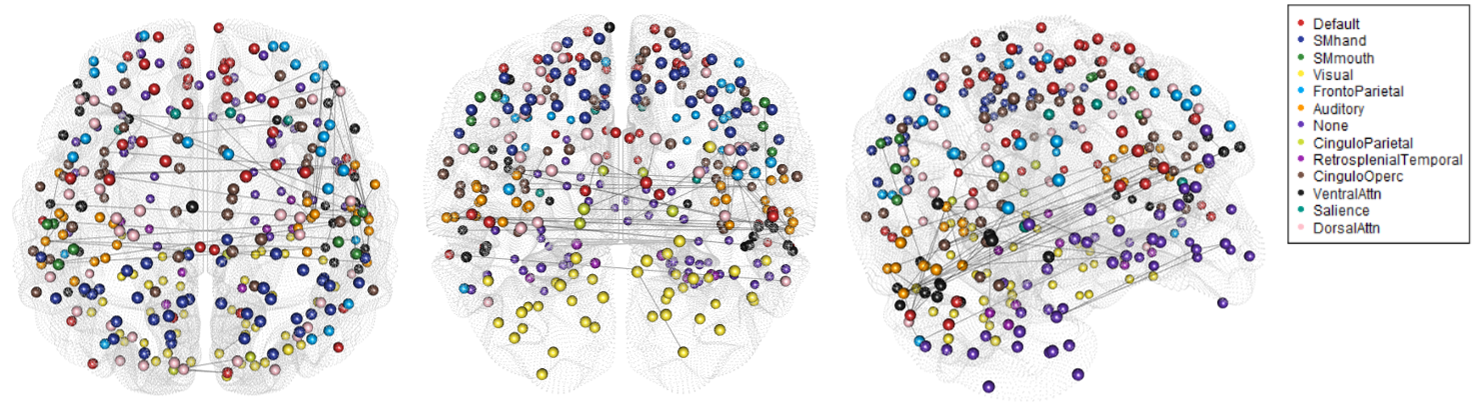}
  \caption{The aggregated network composed across the stationary networks of all subjects. An edge is included in the network if it is estimated $>17$ times (99.9th percentile), based on the sum of edges in adjacency matrices across the stationary networks of all subjects. The aggregate network is shown at three different angles, from the superior in the transverse plane, rostral in the coronal plane, between the coronal and saggitial planes, respectively.}
  \label{fig:aggregatenets}
\end{center}
\end{figure}

We examine the stationary networks between change points. Given we have no \textit{a priori} knowledge of the number of clusters for this data set, we use the cutoff based method to estimate these networks (Section~\ref{section:networkestimation}). We choose a cutoff value which equates to approximately 100 edges. This provides a high level of sparsity in the networks to help interpretability.  Figure \ref{fig:aggregatenets} shows the aggregate network across subjects.  This network is created by first running FaBiSearch on all subjects separately, estimating stationary networks between each pair of change points, and then combining these into one network.  This process involves aggregating information across 161 stationary networks.  An edges is present in the aggregated network if it occurs with a frequency greater than 17 (99.9th percentile).  In the aggregated network, many edges interconnect the left and right hemispheres of the brain.  Additionally, there is some evidence of lateralization of connectivity in the frontal section of the right hemisphere. Furthermore, the nodes which have a high degree are concentrated in the right hemisphere.  Nodes with degree $>5$, $>10$, and $>20$ are in the right hemisphere $51.0\%$, $54.5\%$, and $58.5\%$ of the time, respectively; nodes with more ``hub"-like behaviour are more likely to appear in the right hemisphere.  While reading and language comprehension are distributed processes which involve many regions of the brain \citep{price2012review}, it has been suggested that the right hemisphere is associated with integrating and combining information into higher level processes such as perception and understanding \citep{mashal2005role, st1999semantic} with the more basic processes such as word recognition, syntax, and semantics typically associated with the left hemisphere \citep{binder2003neural, friederici2002towards}.

The five highest degree nodes (in decreasing order, 232, 225, 68, 239, and 226) in the aggregated network (Figure \ref{fig:aggregatenets}) belong to the ventral attention, default, and auditory communities from the Gordon atlas, which play an important role in executing temporally stable operations in the otherwise dynamic reading process.  For example, \cite{seghier2012functional} suggested that the default mode community has various roles during semantic processing, while \cite{corbetta2002control} suggest that components of the ventral attention community are important for ``orienting" attention, especially when presented with unexpected novel or stimuli. The latter suggests that the ventral attention community could be activated by mediating the changes in DFC network structure which are a function of the underlying changes in the story/narrative.  We also see that some nodes are consistently interconnected with a high degree in the auditory community ($p = 24$). This supports the work of \cite{Yao2011} who discovered that the auditory cortex is strongly activated during silent reading, indicative of auditory imagery -- or having an ``inner voice'' while reading. Moreover, the aforementioned high degree nodes are located specifically in Brodmann area (BA) 22, which contains components of Wernicke's area \citep{Binder2170} which has long been associated with language comprehension \citep{bogen1976wernicke}. Thus, it is understandable, from a network perspective, that the nodes related to this area are of high degree and have strong connections to other brain regions during reading. Although reading is a complex task, and people's individual experiences with and understanding of the text play a role in their FC, commonalities do exist. The collection of estimated networks suggest reading is a highly dynamic and integrated process, where more lower level FC features remain stable but higher level processes and patterns are highly variable and different across subjects.

\begin{figure}[ht]
\begin{center}
  \includegraphics[width=0.8\linewidth]{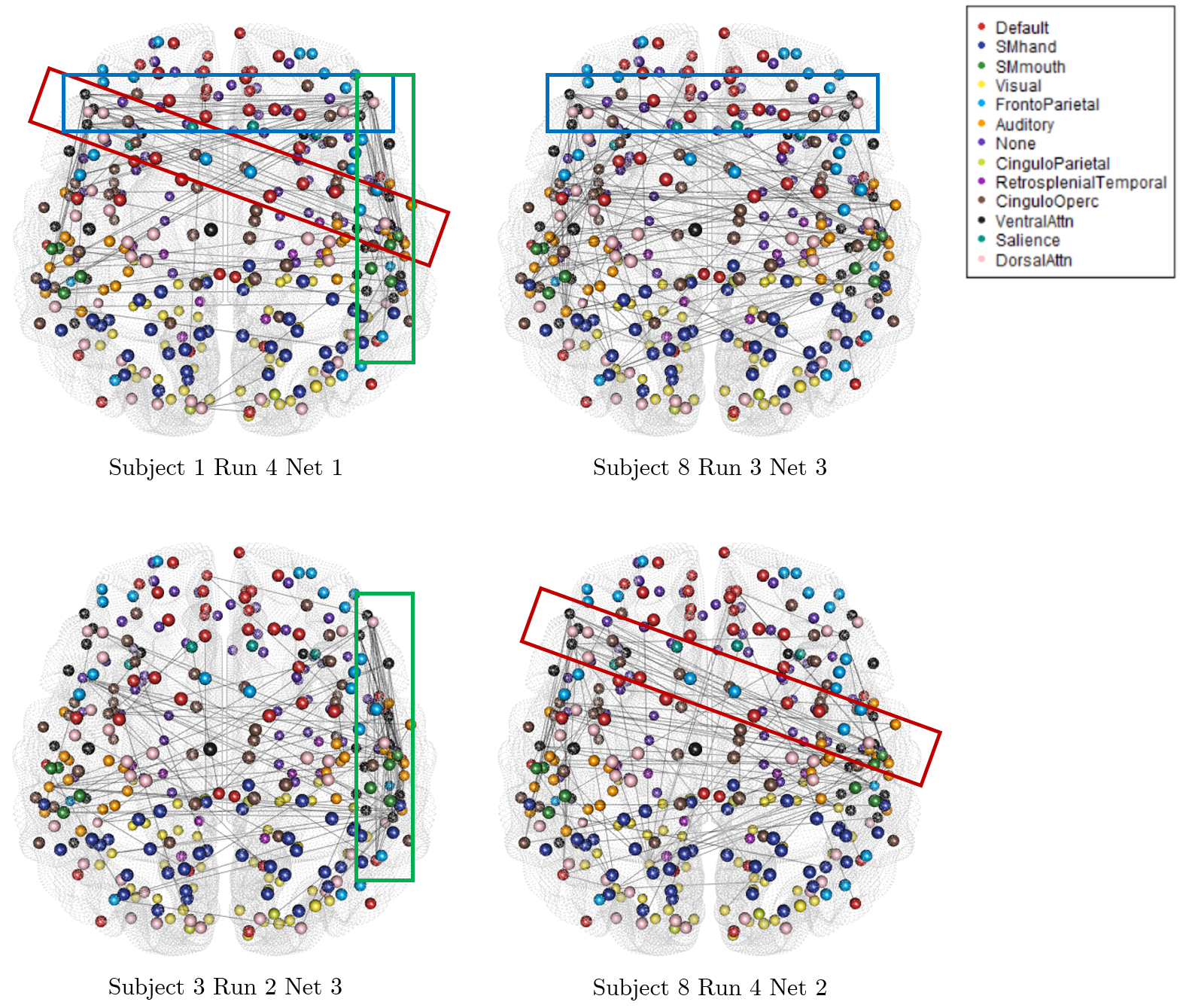}
  \caption{The estimated stationary networks comparing motifs for subject 1 run 4 network 1 (top left), subject 8 run 3 network 3 (top right), subject 3 run 2 network 3 (bottom left), and subject 8 run 4 network 2) for the task based fMRI study.}
  \label{fig:tb_intersimil1}
\end{center}
\end{figure}

We also explore the network characteristics across the stationary networks both within and between subjects. Figure \ref{fig:tb_intersimil1} shows similar network motifs in the first stationary network for subject 1 run 4, and the stationary networks for different subjects and stationary blocks. The first motif, outlined in blue, is a strong cross-midline dependence amongst nodes in the frontal regions of the brain, wherein edges are mostly perpendicular to the midline. The next is somewhat of a pivot, outlined in red, where nodes in the front left area of the cortex are densely connected to nodes in the middle area on the right, close to the temporal region. The last motif, outlined in green, is localized to the right side of the brain, where edges are strongly interconnected all across the outer right side of the cortex. The individual networks are different across subjects, runs, and stationary segments, however the presence of these motifs suggests similarities and stability for network characteristics and FC patterns.

\begin{figure}[ht]
\begin{center}
  \includegraphics[width=0.8\linewidth]{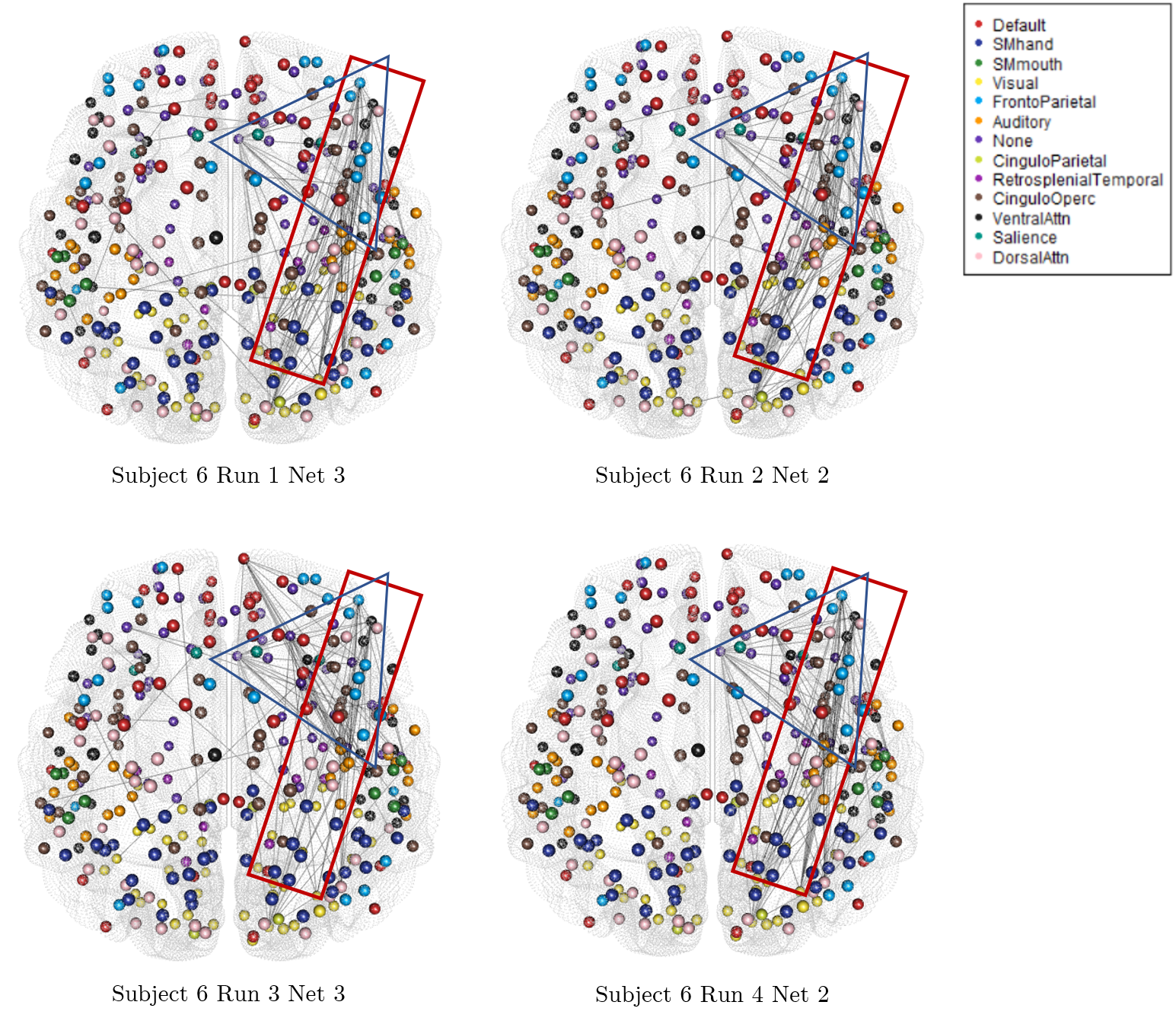}
  \caption{The estimated stationary networks for subject 6 comparing motifs across different runs and stationary networks for the task based fMRI study.}
  \label{fig:tb_intrasimil1}
\end{center}
\end{figure}

Within subjects, there are also some similar patterns across stationary networks. Figure \ref{fig:tb_intrasimil1} shows the similar network features for subject 6 across a variety of different runs and stationary segments. These motifs, or subnetworks, again indicate some stability in the temporal relationships between nodes. These individual networks are still unique, as expected, given in each of these segments the subject was reading a different part of the story, and therefore it is logical that how they mentally process the story differs.

\begin{figure}[ht]
\begin{center}
  \includegraphics[width=0.9\linewidth]{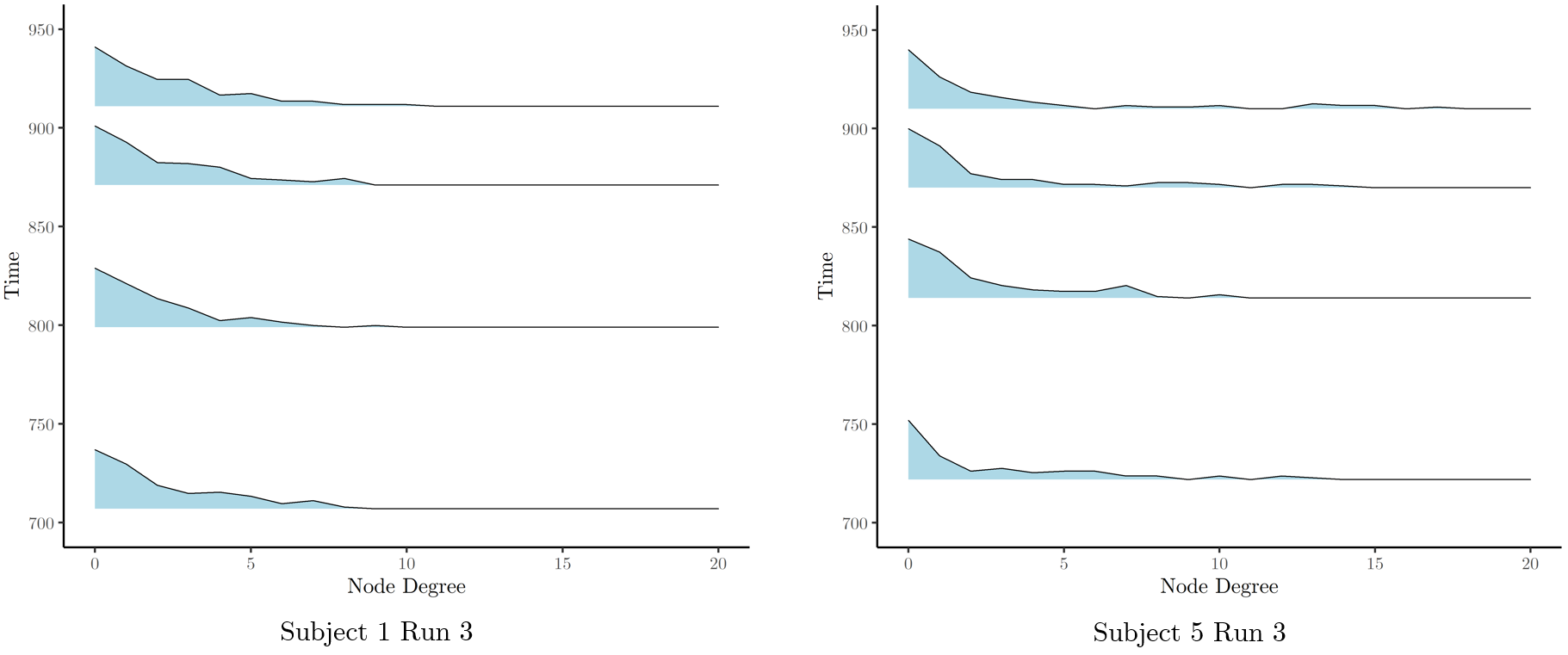}
  \includegraphics[width=0.9\linewidth]{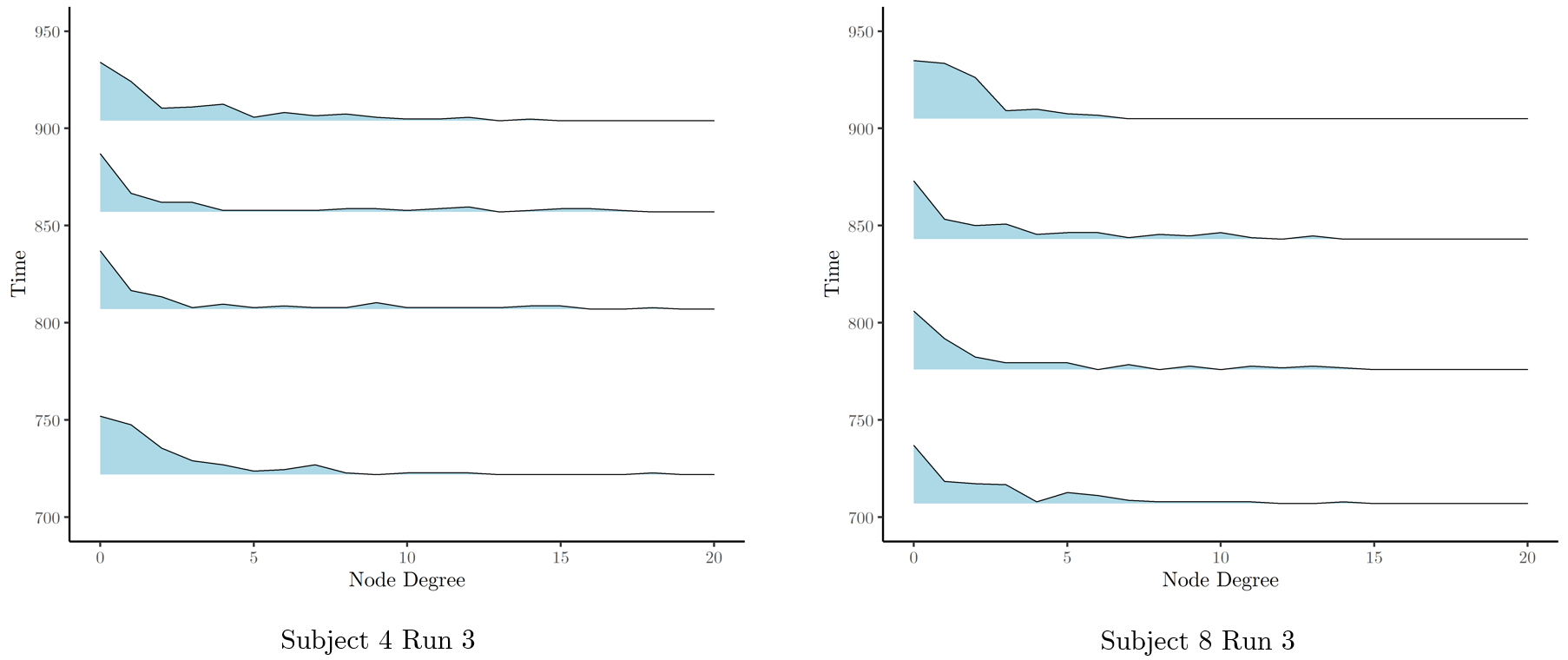}
  \caption{The degree distributions for the  stationary networks in run 3 for subjects 1, 4, 5 and 8 of the task-based fMRI data set. Each degree distribution is centered between detected change points along the y-axis, and the heights are plotted according to a logarithmic scale to highlight the power law.}
  \label{fig:tb_degdistsimil}
\end{center}
\end{figure}

In Figure \ref{fig:tb_degdistsimil} we explore the degree distributions across subjects, specifically, we consider run 3 for subjects 1, 4, 5, and 8.  This run is the most dynamic in terms of story features and also has consistent change points across the chosen subjects. We find that all subjects' individual networks obey the power law of degree distribution, wherein the majority of nodes have low degree and a minority have very high degree. These high degree nodes are perceived as important ``hub" nodes, which is an important and common feature observed in many real-world, including brain, networks. The distributions also seem to follow a similar temporal pattern. For example, for subjects 4 and 8, the distributions appear to move from heavy tailed to lighter tailed from the first stationary network to the next. Then, the second and third stationary networks have nodes with a degree $>10$, indicating greater ``hub"-like behaviour for these nodes. Lastly, the final degree distributions are more concentrated in the $1-10$ degree range. Networks from other subjects and runs also obeyed the power law of degree distribution. However, subjects also vary in the temporal patterns in these distributions. For example, for subjects 1 and 5 (Figure \ref{fig:tb_degdistsimil}), the distributions appear to become more heavily tailed in the last two stationary segments, wherein these segments have node degrees comparatively more concentrated in the $>10$ range. Furthermore, it is possible that the degree distributions and networks themselves may be related to the narrative of the story, or more broadly, the tasks that subjects are participating in  as well as their individual interpretations and perspectives on the story.

%==========================================================================
\section{Discussion}\label{sec:discussion}

\subsection{Extensions}
Non-negative matrix factorization (NMF) is an integral component of FaBiSearch and the proposed graphical method.  Our simulations and data analysis show that NMF is capable of creating a low rank approximation of the data and retain dominant network (or clustering) structures.  For FaBiSearch, we concluded that the algorithms of \cite{NIPS2000_1861} in combination with the generalized Kullback-Leibler divergence performed best for our simulations and fMRI data, however, we also considered other algorithms including the Alternating Least Square (ALS) approach, which minimises an Euclidean-based objective function, that is regularized to favour sparse basis matrices or sparse coefficient matrices. 
Our choices provided a balance between specificity and sensitivity in all signals examined in our large scale simulation study and others not included.  Hence, all have been calibrated on higher dimensions.  The practitioner has the option to use alternatives to these in our R package \textbf{fabisearch} \citep{fabisearch}.

Furthermore, our proposed binary search method in FabiSearch is computationally faster than standard sequential searches, but is possibly limiting in that it is a purely greedy method. In particular, by iteratively halving the search space, binary search improves upon the sequential, exhaustive search of binary segmentation by reducing the worst case search size from $O(T)$ to $O(\log{T})$, where $T$ is the number of indices to be evaluated.  More exploration into non-greedy methods might yield more favourable results and/or decrease computational load. We also intend to explore other change point segmentation methods such as isolate detect \citep{anastasiou}. As an extension to monitoring change points in network objects, we intend to incorporate higher order structures, such as tensors, in the NMF procedure of FaBiSearch.  Finally, we have shown that the residuals between $\boldsymbol{X}$ and $\boldsymbol{W} \cdot \boldsymbol{H}$ in NMF contain important information about the clustering and dependence structure, however, its likely there are alternative methods of extracting this information. 

\subsection{Computation}
We compared FaBiSearch to another method, Network Change Point Detection (NCPD), and found that in the same simulations FaBiSearch has a superior performance across all evaluation criteria. However, in comparison to NCPD, the computational complexity is much greater and thus the time to compute is much greater for FaBiSearch. For example, in Simulation 1, using 48 core machines with 2 Intel Platinum 8260 Cascade Lake at 2.4Ghz and 187GB of memory, FaBiSearch took 25.04 minutes on average while NCPD took 0.75 minutes on average across the 100 iterations.    

\section{Conclusion}\label{sec:conclusion}
In this paper, we characterize and implement a novel multiple change point detection method in the network structure between multivariate high-dimensional time series, called factorized binary search (FaBiSearch), and a method for estimating stationary network structures between detected change points. We assumed the number and location of the change points are unknown a priori. Our methods have several strengths and unique features. Firstly, FaBiSearch scales well to multivariate high-dimensional time series data. This allows us to detect change points in high-dimensional cortical atlas parcellations such as that of \cite{Gordon2016} to characterize whole brain dynamics. There exist very few other change point detection methods capable of handling these wide data sets and we show through simulations how FaBiSearch outperforms one of these methods.

Second, our graph estimation method finds only positive relationships between nodes. Unlike correlation measures, which can take any value in $[-1,1]$, our new method is limited to $[0,1]$. This is because the consensus matrix, which is the basis for this new method, is constrained to $[0,1]$ as it is the arithmetic average of connectivity matrices which are themselves limited to $\{0,1\}$. This makes the corresponding graph more intuitive and understandable, as anticorrelations in a FC context are difficult to interpret.

Lastly, non-negative matrix factorization (NMF) is an integral component of FaBiSearch and the graphical method. This is unique in two distinct ways.  We are, to the best of our knowledge, the first to use NMF for change point detection, network estimation and to apply it specifically to fMRI data. Second, we use NMF to discover dependence structure. While some methods use clustering and/or dimension reduction techniques, these are usually applied in conjunction covariance estimation (or correlation or precision matrices). FaBiSearch solely uses NMF as a method of finding a dependence structure amongst variables, with no intermediary method of finding dependence amongst variables.

We also applied FaBiSearch to a resting-state and a task-based fMRI data set.  For the resting-state experiment, we analyzed the test-retest reliability of dynamic FC, while for the task-based, we explored network dynamics during the reading of Chapter 9 in \textit{Harry Potter and the Sorcerer’s Stone}.  The large scale characterizations of the FC structure has not been explored in these data sets before now.  In general, we detected many change points.  This suggests that regardless of the fMRI study, the FC networks are constantly evolving. We further found common states both across and within subjects in both data sets. This is encouraging as it shows the stability of some features and networks across subjects. Naturally, FaBiSearch could also be used to determine novel biomarkers for neurological phenomena such as disease status or to find network structures corresponding to different thought processes or perceptions based on the subject specific FC.

\section{Acknowledgements}

This research was enabled in part by support provided by WestGrid (www.westgrid.ca) and Compute Canada (www.computecanada.ca).  The first author was supported by the Alexander Graham Bell Canada Graduate Scholarship Master’s (CGS M) award from the Natural Sciences and Engineering Research Council of Canada (NSERC) and the Alberta Innovates Graduate Student Scholarship from Alberta Innovates, Alberta Advanced Education. The third author was supported by the NSERC grant RGPIN-2018-06638 and the Xerox Faculty Fellowship, Alberta School of Business.

\section{Data availability statement}

The authors confirm that the data (for one subject) and code supporting the findings of this study are available within the supplementary materials of the article. The resting-state and task-based data sets were derived from the following public domains \url{http://www.nitrc.org/projects/nyu_trt} and \url{http://www.cs.cmu.edu/~fmri/plosone/}, respectively.

\bibliographystyle{abbrvnat}
\newpage
\bibliography{refs}

\newpage

\section{Appendix}

%%% Tables created using https://tableconvert.com/csv-to-latex %%%

\begin{table}[H]
\begin{center}
    \begin{tabular}{ccccccccc}
    \multicolumn{9}{c}{\large FaBiSearch}\\
    \addlinespace[0.15cm]
        Sim. & mean rank & $\tau$ & TP 1 & FP 1 & TP 10 & FP 10 & Haus. dist. & Compute mins. \\ \hline\hline
        1 & 3.1 & NA & NA & 0.08 & NA & 0.08 & NA & 16.41 \\ \hline
        2 & 5.07 & 100 & 0.85 & 0.18 & 0.99 & 0.04 & 0.0085 & 17.74 \\ \hline
        3 & 6.17 & 100 & 0.39 & 1.57 & 0.75 & 0.72 & 0.1369 & 84.93 \\
        ~ & ~ & 200 & 0.66 & ~ & 0.81 & ~ & ~ & ~ \\ 
        ~ & ~ & 300 & 0.49 & ~ & 0.83 & ~ & ~ & ~ \\ \hline
        4 & 7.15 & 200 & 0.79 & 0.65 & 0.94 & 0.33 & 0.035 & 268.97 \\ 
        ~ & ~ & 400 & 0.79 & ~ & 0.96 & ~ & ~ & ~ \\ \hline
        5 & 5.02 & 100 & 0.78 & 0.56 & 0.93 & 0.25 & 0.0468 & 14.08 \\ 
        ~ & ~ & 200 & 0.78 & ~ & 0.94 & ~ & ~ & ~ \\ \hline
        6 & 6.29 & 100 & 0.47 & 1.61 & 0.96 & 0.69 & 0.0425 & 103.77 \\ 
        ~ & ~ & 200 & 0.51 & ~ & 0.94 & ~ & ~ & ~ \\ \hline
        7 & 5.4 & 100:150 & 0.94 & 0.24 & 1 & 0.13 & 0.1677 & 36.02 \\ \hline
        8 & 5.17 & 100:105 & 0.67 & 0.43 & 0.93 & 0.17 & 0.0615 & 23.49 \\ \hline
    
    \\
    
    \multicolumn{9}{c}{\large NCPD}\\
    \addlinespace[0.15cm]
        Sim. & K & $\tau$ & TP 1 & FP 1 & TP 10 & FP 10 & Haus. dist. & Compute mins. \\ \hline\hline
        1 & 3 & NA & NA & 0.89 & NA & 0.89 & NA & 0.73 \\ \hline
        2 & 3 & 100 & 0.88 & 0.16 & 1 & 0.04 & 0.0063 & 0.39 \\ \hline
        3 & 4 & 100 & 0.23 & 2.98 & 0.57 & 2.12 & 0.3418 & 6.24 \\ 
        ~ & ~ & 200 & 0.05 & ~ & 0.36 & ~ & ~ & ~ \\ 
        ~ & ~ & 300 & 0.41 & ~ & 0.62 & ~ & ~ & ~ \\ \hline
        4 & 3 & 200 & 0.27 & 4.27 & 0.58 & 3.6 & 0.2973 & 31.26 \\ 
        ~ & ~ & 400 & 0.18 & ~ & 0.54 & ~ & ~ & ~ \\ \hline
        5 & 3 & 100 & 0.19 & 2.13 & 0.49 & 1.6 & 0.405 & 1.19 \\ 
        ~ & ~ & 200 & 0.08 & ~ & 0.31 & ~ & ~ & ~ \\ \hline
        6 & 8 & 100 & 0 & 0 & 0 & 0 & NA & 0.57 \\ 
        ~ & ~ & 200 & 0 & ~ & 0 & ~ & ~ & ~ \\ \hline
        7 & 3 & 100:150 & 1 & 0.81 & 1 & 0.79 & 0.1068 & 1.33 \\ \hline
        8 & 2 & 100:105 & 0.05 & 0.8 & 0.16 & 0.69 & 0.339 & 0.71 \\ \hline
    \end{tabular}
\end{center}
\caption{The simulation results for FaBiSearch (top) and NCPD (bottom). Sim., mean rank, $\tau$, TP $i$, FP $i$, Haus. dist., and Compute mins. denote the simulation number, average optimal rank, true change point(s), true positive within $i$ time points, false positives outside $i$ time points, Hausdorff distance within 10 time points, and average compute time in minutes, respectively. $K$ in the NCPD table denotes the number of clusters used for the spectral clustering algorithm.}
\label{tab:simresultstable}
\end{table}

\begin{table}[H]
\begin{center}
    \begin{tabular}{c||cccccccc}
        Sim. & 1 & 2 & 3 & 4 & 5 & 6 & 7 & 8 \\ \hline
        \% Accuracy & 100 & 100 & 99.16 & 99.69 & 99.68 & 96.1 & 98.84 & 99.95 \\
    \end{tabular}
\end{center}
\caption{For each simulation, a comparison of the true networks and the estimated networks using NMF between each pair of change points (stationary blocks).  The percentage was calculated as the proportion of overlap in the off-diagonal elements between the true adjacency matrix and the estimated NMF adjacency matrix.}
\label{tab:SIMgraphingresults}
\end{table}

\end{document}